\documentclass[journal]{vgtc}                

\ifpdf
  \pdfoutput=1\relax                   
  \usepackage{graphicx}                
  \DeclareGraphicsExtensions{.pdf,.png,.jpg,.jpeg} 
\else
  \usepackage{graphicx}                
  \DeclareGraphicsExtensions{.eps}     
\fi%

\graphicspath{{figures/}{pictures/}{images/}{./}} 

\usepackage{microtype}                 
\PassOptionsToPackage{warn}{textcomp}  
\usepackage{textcomp}                  
\usepackage{mathptmx}                  
\usepackage{times}                     
\usepackage{cite}                      
\usepackage{tabu}                      
\usepackage{booktabs}                  

\usepackage{amsmath,amssymb}
\usepackage{listings}
\usepackage{marvosym}
\usepackage{comment}
\usepackage{hyperref,listings,booktabs,overpic}

\usepackage{float}
\usepackage{braket}
\usepackage{subcaption}
\usepackage{enumitem}

\usepackage{color}
\usepackage{url}
\usepackage{soul}

\usepackage{etoolbox}

\newcommand\Tstrut{\rule{0pt}{2.6ex}}         

\newcommand{\x}{{\boldsymbol{q}}}

\newcommand{\X}{{\boldsymbol{p}}}

\newcommand{\Pe}{{\boldsymbol{T}}}

\onlineid{0}

\vgtccategory{Research}
\vgtcpapertype{technique}

\title{Alignment of the Virtual Scene to the Tracking Space of a Mixed Reality Head-Mounted Display}

\author{Ehsan Azimi$^*$, Long Qian$^*$, Nassir Navab, and Peter Kazanzides}
\authorfooter{
\item
 $^*$ Long Qian and Ehsan Azimi contributed equally and are joint first authors.
\item
 Ehsan Azimin, Long Qian and Peter Kazanzides are with Johns Hopkins University. E-mail: \{eazimi1 , long.qian , pkaz\}@jhu.edu.
\item
 Nassir Navab is with Technical University of Munich and Johns Hopkins University. E-mail: nassir.navab@tum.de.
\item
 This work has been submitted to the IEEE for possible publication. Copyright may be transferred without notice, after which this version may no longer be accessible
}

\shortauthortitle{E. Azimi, L. Qian \MakeLowercase{\textit{et al.}}: Alignment}

\abstract{With the mounting global interest for optical see-through head-mounted displays (OST-HMDs) across medical, industrial and entertainment settings, many systems with different capabilities are rapidly entering the market. Despite such variety, they all require display calibration to create a proper mixed reality environment. With the aid of tracking systems, it is possible to register rendered graphics with tracked objects in the real world. We propose a calibration procedure to properly align the coordinate system of a 3D virtual scene that the user sees with that of the tracker. Our method takes a blackbox approach towards the HMD calibration, where the tracker's data is its input and the 3D coordinates of a virtual object in the observer's eye is the output; the objective is thus to find the 3D projection that aligns the virtual content with its real counterpart. In addition, a faster and more intuitive version of this calibration is introduced in which the user simultaneously aligns multiple points of a single virtual 3D object with its real counterpart; this reduces the number of required repetitions in the alignment from 20 to only 4, which leads to a much easier calibration task for the user.
In this paper, both internal (HMD camera) and external tracking systems are studied.
We perform experiments with Microsoft HoloLens, taking advantage of its
self localization and spatial mapping capabilities to eliminate the requirement for line of sight from the HMD to the object or external tracker. The experimental results indicate an accuracy of up to 4 mm in the average reprojection error based on two separate evaluation methods.
We further perform experiments with the internal tracking on the Epson Moverio BT-300 to
demonstrate that the method can provide similar results with other HMDs.
} 

\keywords{Augmented Reality, Mixed Reality, Optical See-Through Head-Mounted Display, Calibration}

\CCScatlist{ 
 \CCScat{K.6.1}{Management of Computing and Information Systems}%
{Project and People Management}{Life Cycle};
 \CCScat{K.7.m}{The Computing Profession}{Miscellaneous}{Ethics}
}

\teaser{
    \centering
    \begin{subfigure}[t]{0.47\linewidth}
        \centering
         \includegraphics[width=\linewidth]{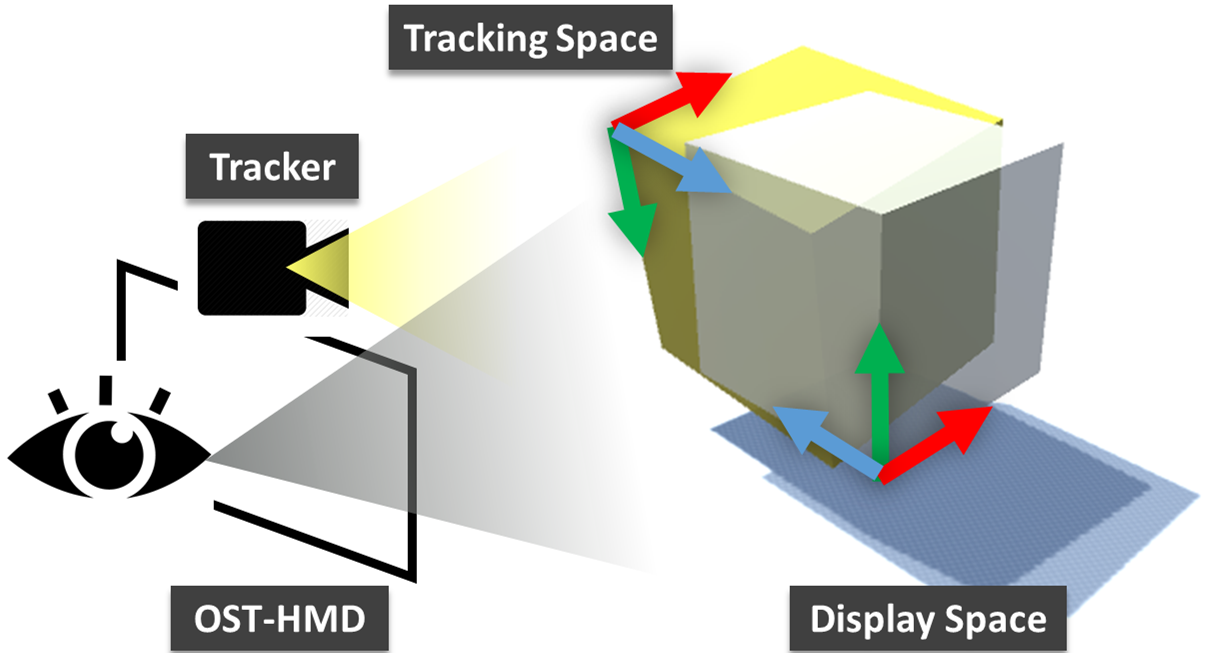}
         \caption{Example with head-anchored tracking system}
    \label{fig:concept_internal}
    \end{subfigure}%
    \hfill
    \centering
    \begin{subfigure}[t]{0.47\linewidth}
        \centering
         \includegraphics[width=\linewidth]{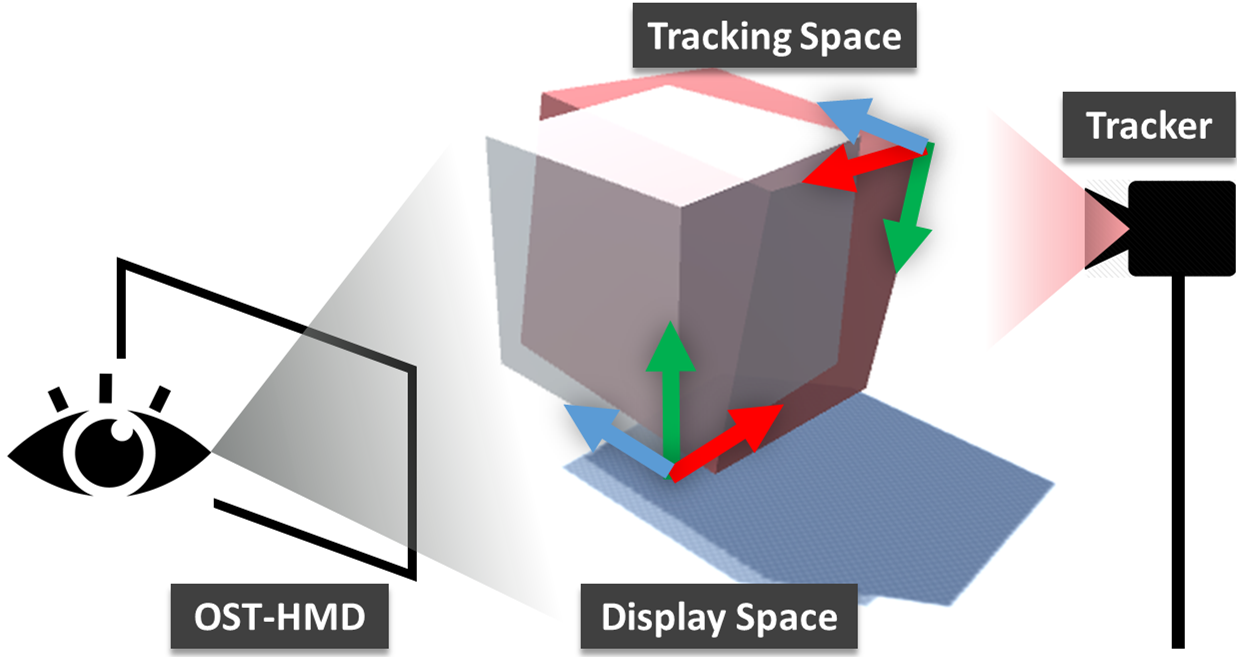}
         \caption{Example with world-anchored tracking system}
    \label{fig:concept_external}
    \end{subfigure}%
    \caption{Problem illustrated: Misalignment between the tracker space and 3D virtual scene; if the OST-HMD is not calibrated with the tracking system, the misalignment between them will cause an unrealistic augmented reality experience.}
    \label{fig:teaser}
}

\renewcommand{\manuscriptnotetxt}{}

\vgtcinsertpkg

\begin{document}


\firstsection{Introduction}

\maketitle
The use of optical see-through head-mounted displays (OST-HMDs) for augmented reality (AR) applications has increased significantly in recent years, due to the engineering advances in commercial OST-HMD devices. One particular area that has remained challenging for these systems is display calibration~\cite{janin1993calibration}. Since augmented reality has to visualize virtual objects in reality, the correct pose and alignment of the displayed objects are of critical importance for the proper user experience using such systems, for example in surgical navigation~\cite{azuma1997survey,azimi2012augmented,sadda2012surgical}. The calibration procedure is aimed to compute the transformation that enables the augmented virtual objects to be represented in the same coordinate system as the real world objects. Proper calibration is also necessary for displaying landmarks in the real world for other applications, including training~\cite{azimi2018ARTrain}.

Many methods intend to improve the display calibration procedure in terms of accuracy~\cite{jun2016calibration,owen2004display,tuceryan2002single}, robustness~\cite{azimi2017robust,moser2016evaluation}, and user-friendliness~\cite{itoh2014interaction,plopski2015corneal,plopski2016automated}.

With the rising commercial interest in recent years, new HMDs such as the HoloLens, Moverio BT-300, Meta Two, and MagicLeap One, have come into the market with a variety of features. They all give users different means to create 3D AR scenes. Despite their differences, they all share the need for a display calibration to create a proper mixed reality experience. Also, depending on the platform the user may or may not have access to the projection matrices. This access, however, is limited to the device dependent transformation from the display to some sort of tracking coordinates, and cannot provide the transformation from the display to the eye; therefore, the transformation chain to the user's eye is incomplete.

With the HoloLens, even though its spatial mapping and integrated IMU sensing enables the user to create stable virtual objects in the real world, there has not yet been a systematic work to properly align the created virtual content with the world in a dynamic setting, which is required in many augmented reality applications. Although the Vuforia SDK~\footnote{Vuforia SDK:\url{https://vuforia.com/}} has implemented image-tracking support on the HoloLens, it does not completely address the alignment between virtual content and the user's view. 

\begin{figure}[t]
    \centering
    \includegraphics[width=0.99\linewidth]{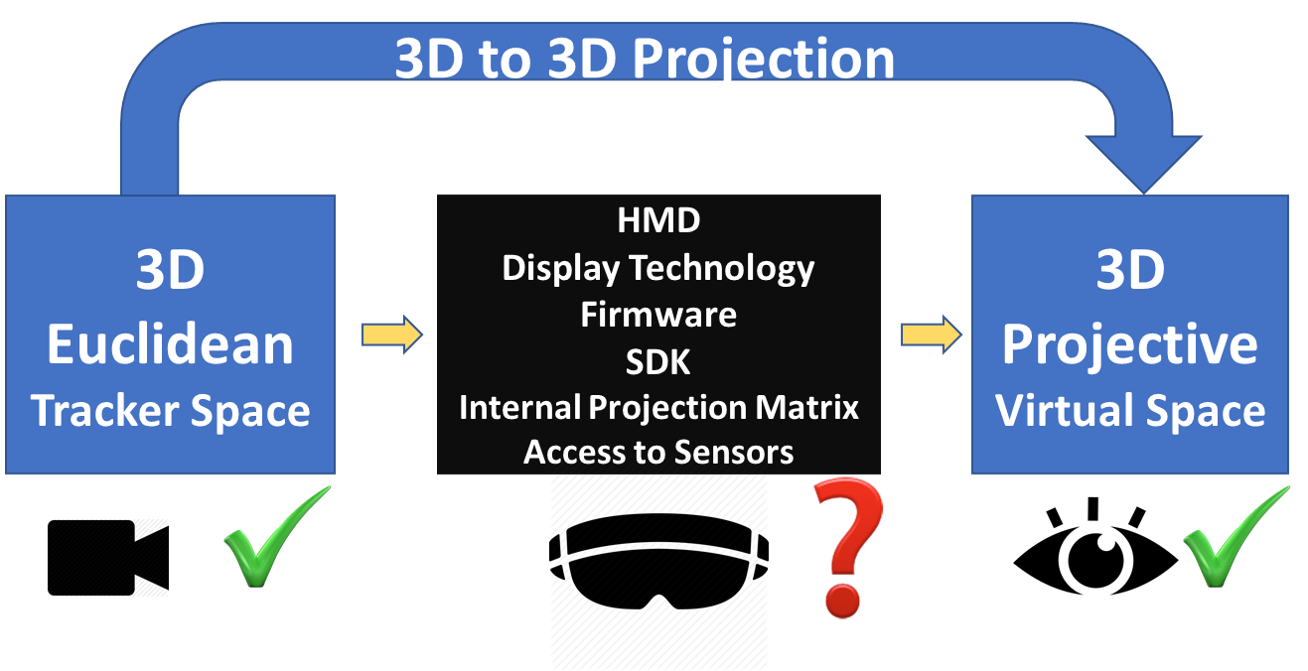}
    \caption{Conceptual schematic of blackbox approach towards display calibration: regardless of the internal features of an HMD, as long as there is access to the generated 3D scene for the user and the tracking data, the display can be calibrated with the proposed method.}
    \label{fig:blackbox}
\end{figure}

Our approach considers any designated OST-HMD as a blackbox, and treats the data from a tracking system as the input and the visualization of a virtual 3D object in the eyes of the observer as the output. It then proposes a procedure to identify this blackbox by computing the 3D to 3D projection matrix that corrects the misalignments of the real object with its virtual counterpart in the user's eyes. Regardless of the intermediate processes in creating the virtual scene, this final alignment in the 3D perceived scene is what matters for the observer and affects the mixed reality experience. The concept of such a blackbox is represented in Fig~\ref{fig:blackbox};
we demonstrated our method on the HoloLens and Moverio BT-300, but it can be easily applied to other OST-HMDs.

Another challenge with OST-HMD calibration is the need for several steps of alignment and the fatigue and inaccuracies caused by such repetition. Also, the calibration is no longer valid when the user changes or the HMD slips relative to the user's eyes. In an effort to make the process of calibration faster and more intuitive with fewer alignment steps, we provide an extension to our work where the user aligns multiple points of a 3D object at the same time.
This is similar to the MPAAM method proposed by Grubert et al.~\cite{grubert2010comparative}, but with a simpler setup because it uses alignments of multiple corners of a single 3D object.

The contributions of the present study can be summarized as:
\begin{itemize}[noitemsep]
\setlength{\itemsep}{2pt}
\item An end to end solution to find the 3D to 3D projection of a given OST-HMD that aligns the virtual objects in the 3D scene with their real counterparts.
\item A multipoint alignment version of our approach that is more intuitive and makes the display calibration considerably faster and easier.
\item Experimental verification of the methods, using different setups and tracking systems, geometrical models, and multimodal evaluation procedure.
\item Use of HMD spatial mapping for calibration and tracking, thereby eliminating the line of sight requirement from the HMD to the tracker or to the tracked object. 
\item A novel evaluation method that objectively assesses the calibration accuracy.
\end{itemize}

In the following sections, we will first explain some of the existing methods in the literature. Next, we derive the necessary equations for the calibration in such setups. Then, we describe the implementation of our calibration process. We deploy our calibration application to both Microsoft HoloLens and Moverio BT-300 to demonstrate that any arbitrary application that displays 3D geometry on the screen can use the proposed method to perform the calibration. We evaluate our new method in two different ways, analyze the errors and validate its accuracy.

\section{Background}

\subsection{OST-HMD Calibration}

With an optical see-through head-mounted display (OST-HMD) there is no direct access to the view of the user. Therefore, a user-dependent calibration procedure is generally required in order to correctly align the virtual content with its real counterpart. The Single Point Active Alignment Method (SPAAM)~\cite{kato1999marker,tuceryan2002single} is one of the widely applied methods to perform such display calibration due to its simplicity and accuracy. In SPAAM, 2D-3D point correspondences are collected by the user, and then a mapping from the 3D point cloud to its 2D screen coordinates is calculated using Direct Linear Transform (DLT)~\cite{hartley2003multiple}. Methods have been proposed to improve the user interaction~\cite{genc2002practical,qian2016reduction} and robustness~\cite{azimi2017robust,makibuchi2013vision} based on SPAAM. Unlike modeling the mapping from 3D point sets to the 2D screen coordinates as a projection, the DRC method~\cite{owen2004display} takes the physical model of the optics into account. Recently, interaction-free display calibration was developed with the help of an additional eye-tracking system~\cite{itoh2014interaction,plopski2015corneal}.

These methods calibrate one screen of the OST-HMD and one eye of the user. It is possible to perform the calibration procedure twice, in order for the rendered graphics on a stereoscopic OST-HMD to correctly align with the user's two eyes. Stereo-SPAAM~\cite{genc2000optical} is a variant of SPAAM that simultaneously calibrates both eyes with a stereoscopic OST-HMD. Simultaneous calibration of a stereoscopic OST-HMD is better conditioned by adding the physical constraints of two eyes, such as interpupillary distance (IPD). However, Stereo-SPAAM finds the projection matrix from the virtual camera formed by the eye to the planar screen; it merely calculates two separate 2D-3D mappings with the same underlying model.

For all of these methods to work, there is a need to access and modify each screen's projection matrix.
That, however, may not be possible in some setups where the manufacturer may only provide access to the final 3D visualization of virtual content. In fact, most OST-HMDs enable the user to create 3D visualization of virtual content in front of the users' eyes. Therefore, regardless of the level of access to internal settings of individual screens of the OST-HMD, it is possible to use the 3D representation of virtual content, which is the final output in such systems and easily accessible for the user. Therefore, from this perspective the projection is from the 3D world to a 3D space rather than two planar screens. In this scenario, the representation of information can be thought of as three-dimensional in space instead of two-dimensional within the screen coordinate system. As a consequence, the mapping model becomes a 3D-3D registration procedure.

We should emphasize that while there may be different interpretations of display calibration, in our study, the term ``display calibration'' pertains to computing the 3D to 3D transformation matrix that corrects the misalignments of the real object with its virtual counterpart in the user's eyes, so that the augmented virtual objects can be represented in the same coordinate system as the real world objects.
The calibration procedure, in a general sense, is aimed to compute the transformation that enables the augmented virtual objects to be represented in the same coordinate system as the real world objects. Any process that helps with this can be considered to be part of the display calibration.
For instance, what is referred to as calibration in the HoloLens is in fact IPD measurement, but similarly to our method the end result would adjust the transformation that is applied to the CGI, as explained in Section ~\ref{sec:method_geo}.
 
\subsection{Tracking Systems}

For a proper augmented reality application, virtuality should be perceptually registered with real objects during motion~\cite{azuma1997survey}, so that the rendered graphics reflect the motion. Augmented reality based on head-mounted displays usually incorporates two categories of tracking system: head-anchored tracking system (also called ``inside-out'')
and world-anchored tracking system (also called ``outside-in'')~\cite{wang2017prioritization}.

Many head-mounted display systems include an embedded front-facing camera, which can serve as the optical tracker for AR applications. A head-anchored tracking system has the advantage of providing similar line of sight to that of the user, but its performance is limited by the size, power consumption, and computational cost. Calibration of an OST-HMD using a head-mounted camera has been proposed~\cite{fuhrmann2000practical,genc2001optical,kato1999marker}. Marker-based tracking algorithms have the advantage in simplicity and robustness~\cite{garrido2014automatic,kato1999marker}, and marker-free tracking algorithms offer better user experience~\cite{joseph2015versatile,newcombe2011kinectfusion}.

In a world-anchored tracking system, the pose of the tracker coordinate system remains unchanged with respect to the world coordinate system. Without the constraints imposed by power, computational resources and the type of technology used, a world-anchored tracking system can potentially be more accurate. Examples of world-anchored tracking systems include reflective markers tracker, electromagnetic sensing, and projective light-based tracker.

In this paper, calibration of an OST-HMD based on a head-anchored tracking system and a world-anchored tracking system are both studied and presented.

\section{Method}
We first explain the method we use to solve the transformations using different linear geometric models in Section~\ref{sec:method_geo}. Then, in Section ~\ref{sec:method_multi} we describe a multipoint version of our approach which makes the calibration procedure faster and more intuitive. 

\label{sec:method}

\subsection{Calibration with blackbox approach}
\label{sec:method_geo}
In this section, we look into the OST-HMD calibration and projection matrix of a 3D display space for the derivation of the required transformation. 

We have a transformation $\Pe(\cdot)$ which maps 3D points from the world coordinates to a 3D virtual environment. Basically, if we are given the points $\x_1,\cdots,\x_n$, through the transform we observe $\X_1,\cdots,\X_n$ such that
\begin{equation}
\X_i = \Pe(\x_i) \quad i =1,\cdots,n.
\label{equ:one}
\end{equation}

We assume that both $\X_i$ and $\x_i\in \mathbb{R}^{3}$. The goal is to estimate $\Pe$ based on a set of observations in the form of $(\x_i,\X_i)$ for $i=1,\cdots,n$. More specifically, the measurement of $\x_i$ is obtained from the tracking system, while the information of $\X_i$ is pre-defined and visualized on the OST-HMD. With the calculated transform $\Pe(\cdot)$, a point from the tracker coordinate system is mapped to that of the display coordinate system.

We further assume that the transformation $\Pe(\cdot)$ is linear, and since our aim is to find the transformation between the 3D sensor tracking coordinate system and the 3D scene camera (visualization) coordinate system we assume that it is an affine transformation (12 unknown parameters), as the transformation between coordinate systems is affine.
To verify this assumption, we also solve for the general case where the transformation $\Pe$ is a perspective transformation, with
15 unknown parameters (excluding an arbitrary scale parameter).
In addition, because fewer unknown parameters require fewer calibration alignments and thus can considerably reduce the burden on the user, we consider an isometric transformation that has 6 unknown parameters. 

To better understand our blackbox approach, a brief mathematical overview is provided.
The mathematical representation of a 3D-3D transformation is:
\begin{equation}
\hat{\X_i} = \begin{bmatrix}
\: T \:
\end{bmatrix}_{4\times 4} \cdot \hat{\x_i}, \quad
\label{equ:two}
\end{equation}

To solve the calibration problem, we look into the projection $T$ that minimizes the reprojection error: 
\[\min_T E_{reproj}
\]

Different methods for solving these transformations have been studied~\cite{newman2004ubiquitous,horn1987closed,umeyama1991least,moser2016towards,moser2016evaluation}.
Both perspective transformation and affine transformation are calculated with the Direct Linear Transformation (DLT) algorithm, with the objective of minimizing total algebraic error~\cite{hartley2003multiple}. For an isometric transformation, the problem is equal to registration of two rigid 3D point sets; therefore, the absolute orientation method of Arun is used~\cite{arun1987least}, with the objective of minimizing least-square error of the registration.

The HMD display generates the 3D image by presenting two 2D perspective images, one for each eye, of a 3D scene from two slightly different viewing positions. Each one has its own projection matrix, $P$. We further assume that each HMD comes with its default configuration and has a preset internal projection matrix which is part of the ``blackbox'' and can be described as:
$$
\text{Left Eye Default:}  ~\begin{bmatrix}
\: P_{LD} \:
\end{bmatrix}_{3\times 4} ,  
$$
\vspace{-12pt}
$$
\text{Right Eye Default:}
\begin{bmatrix}
\: P_{RD} \:
\end{bmatrix}_{3\times 4}  
$$

In the proposed method, the computed transformation, $T$, is in fact applied to both of these internal projections, which results in the following effective projection matrices for the left and right eye, $P_{LE}$ and $P_{RE}$:
$$
\begin{bmatrix}
\: P_{LE} \:
\end{bmatrix}_{3\times 4} = \begin{bmatrix}
\: P_{LD} \:
\end{bmatrix}_{3\times 4} \cdot \begin{bmatrix}
\: T \:
\end{bmatrix}_{4\times 4} \quad
$$
$$
\begin{bmatrix}
\: P_{RE} \:
\end{bmatrix}_{3\times 4} = \begin{bmatrix}
\: P_{RD} \:
\end{bmatrix}_{3\times 4} \cdot \begin{bmatrix}
\: T \:
\end{bmatrix}_{4\times 4} \quad
$$

These are the corrected projection matrices. Therefore, the computed transformation, in effect, adjusts the default internal projection matrices to correct the misalignments in visualizing the virtual objects with respect to the real scene.
In other words, the computed elements in this matrix will adjust the original (blackboxed) display calibration, e.g., aspect ratio, focal length, and extrinsic transformation.
In general, a $3 \times 4$ projection matrix contains 11 degrees of freedom (6 for the camera extrinsics and 5 for the camera intrinsics). For stereo visualization, one common approach is to use the same projection matrix for both eyes, except with a translation (obtained using the IPD) along one coordinate direction, for a total of 12 degrees of freedom. We consider different types of transformations (e.g., isometric, affine, and perspective), which differ in the number of degrees of freedom and therefore also in the extent to which they can adjust the blackbox projection matrices.

We analyze the error that results from the use of each of these transformations to determine which one better represents the correct model.
As will be discussed in Section~\ref{traintest}, the error analysis is performed with a different set of data that was not used for calibration.

The alignment task is performed by humans and is therefore prone to errors. The RANSAC~\cite{fischler1981random} algorithm is used to find the most accurate transformation and reject outliers based on the reprojection error of the samples:
\[E_{reproj}=\dfrac{\sum_{i=1}^n  \sqrt{(\X_i-\Pe(\x_i))^2}}{n}
\]

\begin{figure}[!tb]
  \centering
  \includegraphics[width=0.35\linewidth]{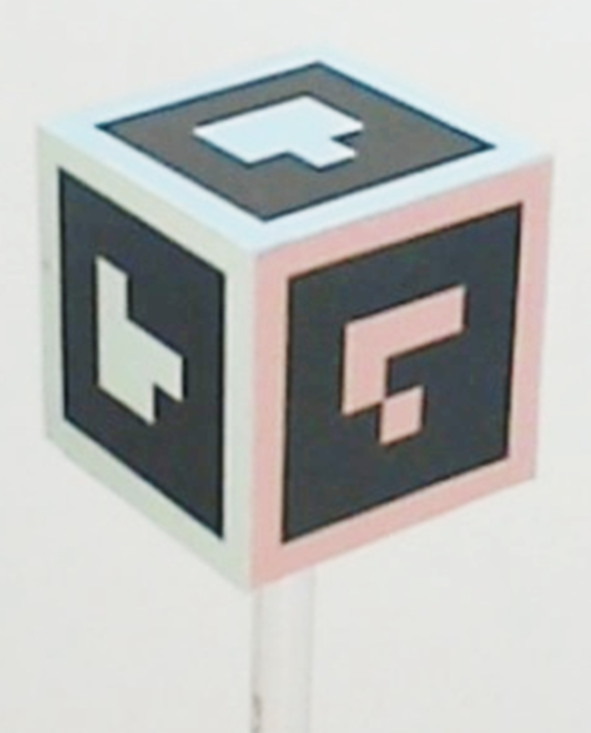}
  ~
  \includegraphics[width=0.45\linewidth]{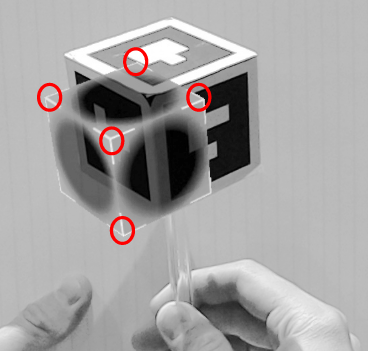}
  \caption{Left: tracked object; Right: Multipoint single object alignment procedure.}
  \label{cube_multipoint}
\end{figure}

The following process is proposed for computing the transformation from the real to the virtual scene coordinate system. In this point alignment scheme, the user has to align points from the real world observed by the tracking system $\x_i$, with that of the virtual 3D world $\X_i$.
In Stereo-SPAAM, the alignment target is designed to be a virtual disk displayed with some disparity on two separate screens~\cite{genc2001optical}. However in this setup, to utilize the unique depth cue characteristics of the 3D visualization, we used a $2''\times2''\times2''$ cube with different fiducial markers attached on its faces and painted each face with a different color that matches its counterpart in the virtual cube, as shown in Fig.~\ref{cube_multipoint}-left.
For each calibration, the user is asked to perform multiple alignments in which (s)he aligns a specified corner of a real cube with the corresponding corner of the virtual cube in the OST-HMD display.

\subsection{Multipoint Alignment Using a Single 3D object}
\label{sec:method_multi}
The calibration procedure described in the previous section is a tedious task that requires the user to perform a relatively large number of alignments (20, in our implementation). Not only do these repetitions increase the time of calibration, but they also reduce the accuracy because the task is performed by humans~\cite{qian2016reduction}.

In an extension to our method, we developed a calibration technique that requires the user to align multiple corners of a real 3D object with its virtual counterpart in the scene, which leads to an easier and faster calibration process.
This approach is feasible because (as will be shown later) the isometric transformation, which preserves the distances between points, produces sufficiently accurate calibration results.
An illustration of the multipoint alignment procedure is presented in Fig.~\ref{cube_multipoint}-right. As it is shown, the user has to align the 5 corners of the real cube with their corresponding virtual counterparts in the scene.
Of course in theory, 3 non-collinear points in space are enough to fully determine the pose of a 3D object, but we use 5 points to make the alignment easier, provide a better depth cue, and collect additional points.
With this method, 4 alignments produce 20 point correspondences, which then go through the same RANSAC and DLT or least square method to compute the corresponding transformations. 

It should be considered that even though multiple points are used to make the calibration more user-friendly and efficient, our approach differs from MPAAM because (1) we use multiple points on a single 3D object and (2) the alignments are from 3D tracker space to 3D virtual space and therefore it follows the 3D-3D formulation. Thus, since our proposed single point model is different from SPAAM, the extended multipoint version is also different from MPAAM.

\section{Implementation}

We used Microsoft HoloLens (Microsoft Inc., USA) and Moverio BT-300 (Epson Inc., USA) for our experiments. We used two different HMDs to show that the proposed method is generic and not limited to any specific setup.
Due to the different characteristics of different tracking systems and to verify the applicability of our method, both head-anchored and world-anchored tracking systems were studied for the calibration.

It should be clarified that our method does not rely on specific features of the HoloLens. First, we do not require the HoloLens spatial mapping for either tracking scenario. For the world-anchored tracking system, we
take advantage of it to obtain the pose of the HMD with respect to the external tracker, but could instead attach markers to the HMD and directly track it.
Second, the HoloLens embedded calibration procedure measures the IPD of the user and then adjusts the pose of the virtual content based on the measured IPD. However, in our experiments we do not adjust the default IPD for each subject. Therefore, our method does not rely on those internal settings to compute the correct projection.

\subsection{Calibration with Head-Anchored Tracking System}
\label{sec:head-anchored}

\begin{figure}[t]
    \centering
    \includegraphics[width=0.65\linewidth]{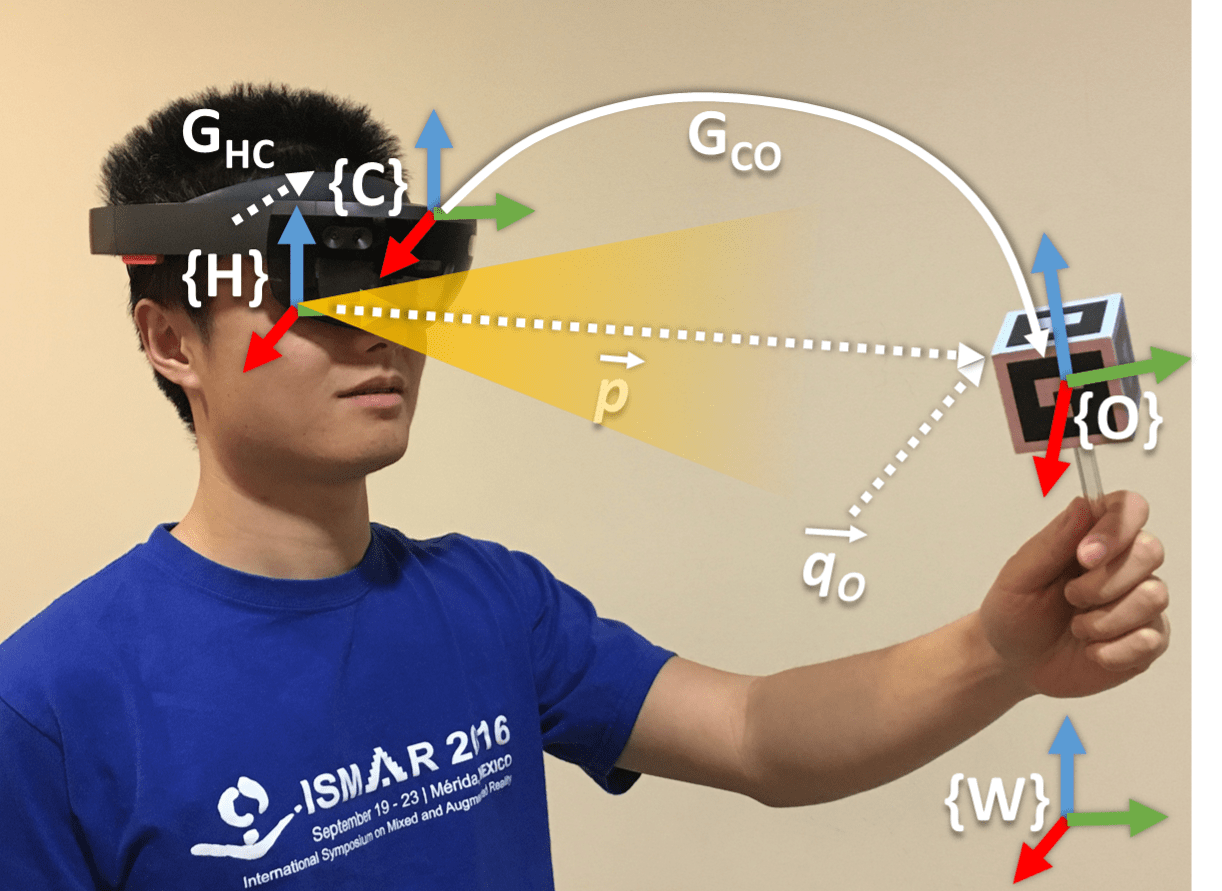}
    \caption{Implementation of the calibration with head-anchored tracking system. The HoloLens embedded RGB camera is used as the head-anchored tracker. The transformation map of the calibration procedure is demonstrated.}
    \label{setup_internal}
\end{figure}

\noindent\textit{i) Microsoft HoloLens:}
In this case, the embedded HoloLens front-facing RGB camera is used as the head-anchored  tracker. Fiducial markers are attached to a real object that is held by the user. The coordinate systems of the tracker, object and HMD display are represented as $\{C\}$, $\{O\}$ and $\{H\}$, respectively, as shown in Fig.~\ref{setup_internal}. Since the camera is rigidly mounted on the HMD, the extrinsic geometric transformation between the camera and the HoloLens $G_{HC}$ is fixed. The point for alignment is fixed at $\vec{q_O}$ with respect to the coordinate system of $\{O\}$. Its corresponding virtual point is at $\vec{p}$ in the HMD display coordinate system $\{H\}$.
The pose of the tracked object $G_{CO}$ is determined with a marker-tracking package HoloLensARToolKit~\footnote{HoloLensARToolKit: \url{https://github.com/qian256/HoloLensARToolKit}} at runtime. 
Eventually, the point sets $\{\x\, |\, \x_i = G_{CO,i}\cdot\vec{q_O},\, i=1,\cdots,n \}$ and $\{\X_i\,|\,i=1,\cdots,n\}$ are used for the OST-HMD calibration described in Section~\ref{sec:method}.
\\
\\
{\noindent\textit{ii) Moverio BT-300:} 
Similar to the HoloLens, we deployed our application to Moverio BT-300 and used its embedded front-facing RGB camera for tracking. The user holds the same object with attached ARTags for tracking.
Moreover, we treated the BT-300 as a blackbox and did not perform standard display calibration (e.g., using SPAAM to estimate a 3x4 projection matrix for one eye and a measured IPD to find the projection for the other eye) before applying our method. We intentionally used a projection matrix and IPD that caused poor alignment of the virtual objects to better demonstrate the capability of our calibration method.}

\subsection{Calibration with World-Anchored Tracking System}

\begin{figure}[t]
    \begin{subfigure}[t]{0.525\linewidth}
        \centering
        \includegraphics[width=0.9\linewidth]{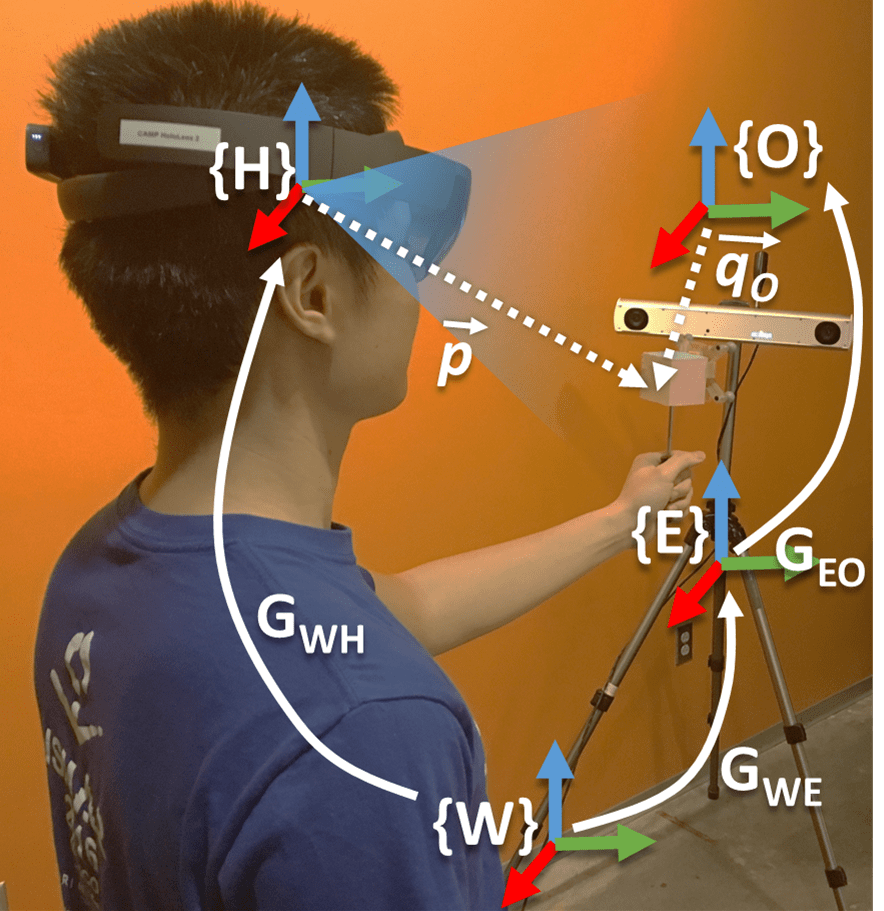}
        \caption{Transformation graph}
        \label{setup_external}
    \end{subfigure}%
    \hfill     
    \begin{subfigure}[t]{0.475\linewidth}
        \centering
        \includegraphics[width=0.9\linewidth]{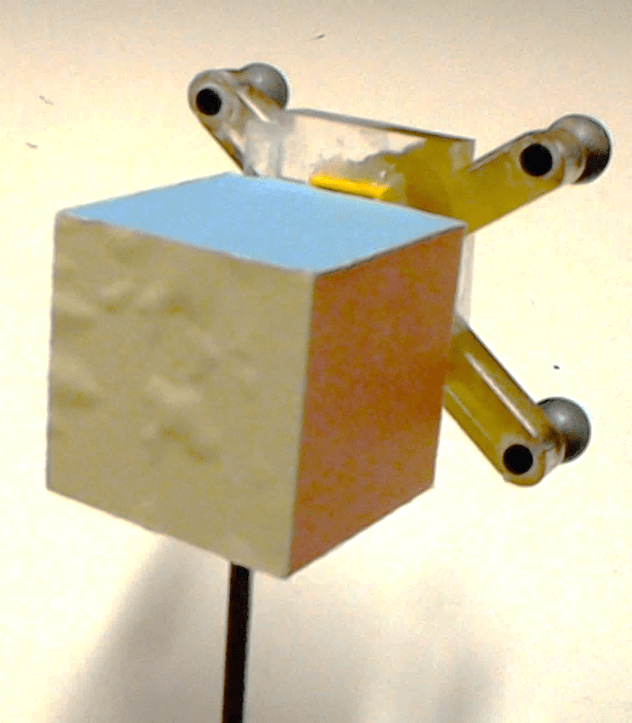}
        \caption{Tracked object}
        \label{cube_external}
    \end{subfigure}
    \caption{Implementation of the calibration with world-anchored tracking system (fusionTrack 500). Passive spherical markers form a frame that is attached to the colored cube for tracking.}
\end{figure}

The coordinate systems of the tracker, object, HMD display and world are represented as $\{E\}$, $\{O\}$, $\{H\}$ and $\{W\}$, respectively, as shown in Fig.~\ref{setup_external}.   
It should be noted that both $\{C\}$ and $\{E\}$ represent the tracker coordinate system. However, since our general workflow is different for these two hardware systems, we refer to the HMD camera as $\{C\}$ and external tracker as $\{E\}$ to reduce ambiguity when explaining both methods. The main conceptual difference between head-anchored and world-anchored tracking systems for our calibration process is as follows.

In the head-mounted tracker case, the transformation $G_{HC}$ between the tracker $\{C\}$ and the HMD display $\{H\}$ is fixed, but this is not the case for the world-anchored tracker, where the transformation $G_{HE}$ is expressed as $G_{HE} = {G_{WH}}^{-1} \cdot G_{WE}$. Since the world-anchored tracker does not change its pose in the room, $G_{WE}$ is fixed. Therefore, an extra component is needed to maintain and update the transformation $G_{WH}$ between the world and HMD display $\{H\}$, so that the transformation between the tracker and the display $G_{HE}$ can be determined. The SLAM-based spatial mapping capability of the HoloLens fills this gap and completes our transformation chain from the tracked object to the user's view.
In other words, the transformation from the tracker to the object is known because it is tracked and the external tracker is fixed in the world. Spatial mapping provides and updates the pose of the HMD with respect to the world and therefore we can close the transformation chain and find the pose of the object relative to the HMD. Therefore, direct line of sight between the HMD camera and the object is not needed, as long as the user (not the HMD) can see the object to perform the alignment and the tracker remains in the same position in the world.
After calibration, direct line of sight is not needed.
Practically, the HMD spatial mapping will not rely on the object, but rather on self-localizing within the room environment (e.g., with respect to large features such as walls). To the best of our knowledge, this is the first time that the self localization and spatial mapping of a HMD is used for the calibration.

For HMDs that do not perform spatial mapping, an alternative method would be required, such as using the external tracking system and mounting fiducial markers to also track the HMD.

In our implementation for the world-anchored tracker experiment, we used a fusionTrack 500 (Atracsys LLC, Puidoux, Switzerland). Passive spherical markers compose a frame which is attached to the cube that is used for alignment  (Fig.~\ref{cube_external}). A component is written in C++ to handle processing and transmission for the external tracker. Tracking information of the tracked objects is obtained with a frequency greater than 300 Hz and is transmitted to the HoloLens via a wireless network. The UDP protocol was used to minimize latency and allow real-time performance.

\section{Experiments}

\begin{figure*}[tbh]
  \centering
    \includegraphics[width=0.86\linewidth]{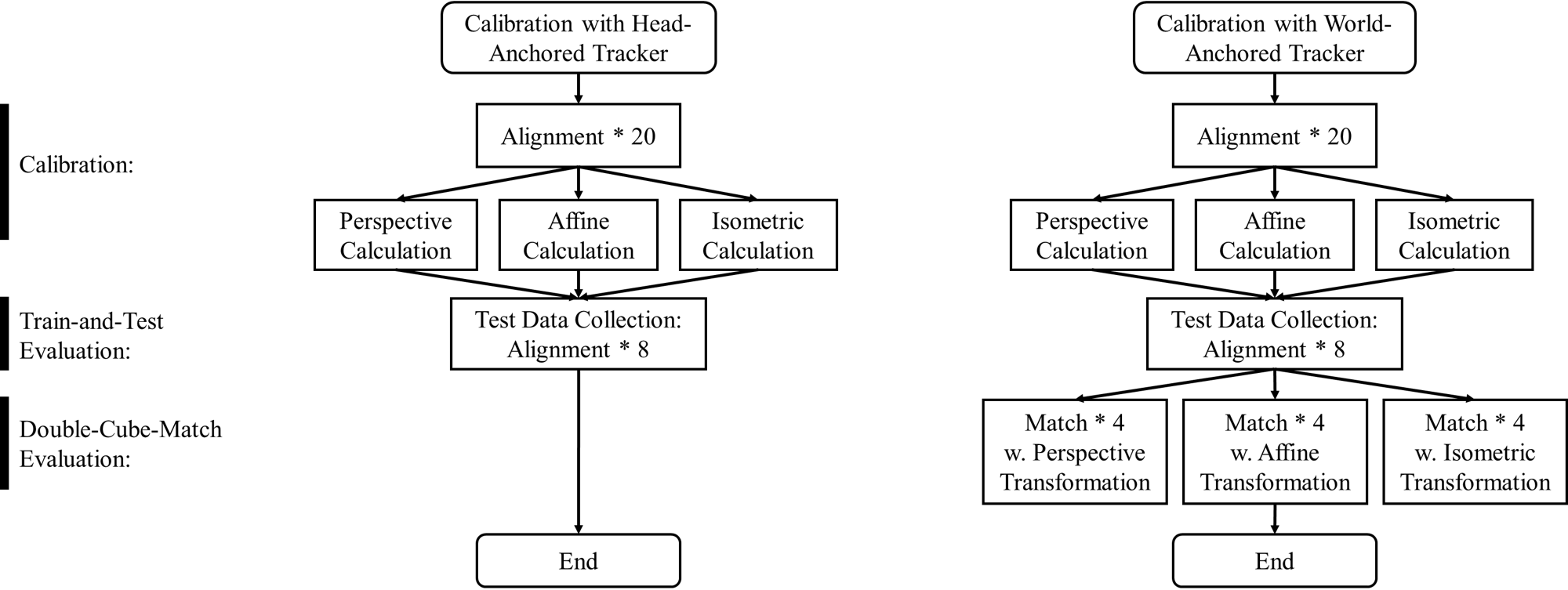}
    \caption{The workflow of calibration with head-anchored tracker and world-anchored tracker}
    \label{workflow}
\end{figure*}

To analyze and evaluate our proposed calibration method, experiments were carried out with the 
head-anchored and world-anchored trackers, with setups shown in
Fig.~\ref{setup_internal} and Fig.~\ref{setup_external}, respectively.
For the world-anchored tracker case, we attached passive markers with reflective spheres to the painted cube (Fig.~\ref{cube_external}) and since the alignment is done relative to the cube, a pivot calibration was performed to obtain the transformation from the tracked passive markers to the attached cube. 
The calibration workflow diagrams for the two trackers are depicted in Fig.~\ref{workflow}.

\begin{figure*}[tbh]
    \begin{subfigure}[t]{0.3\linewidth}
        \centering
        \includegraphics[width=0.95\linewidth]{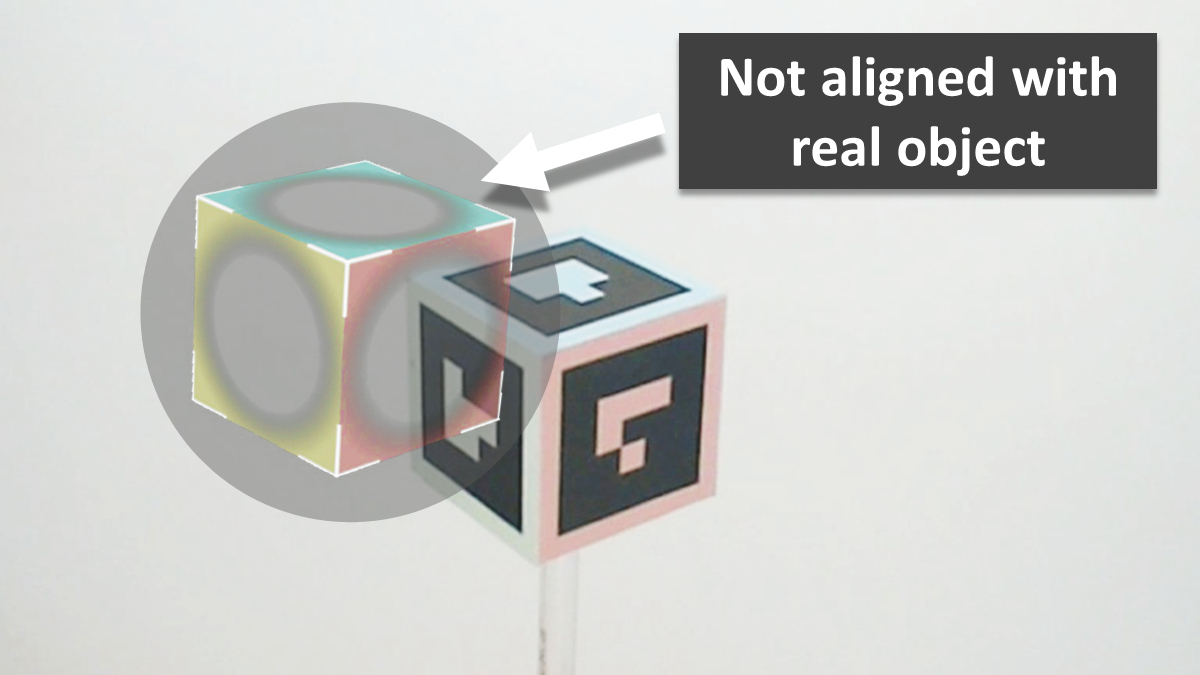}
        \caption{See-through view before calibration}
        \label{calib_internal_before}
    \end{subfigure}%
    \hfill
    \begin{subfigure}[t]{0.3\linewidth}
        \centering
        \includegraphics[width=0.95\linewidth]{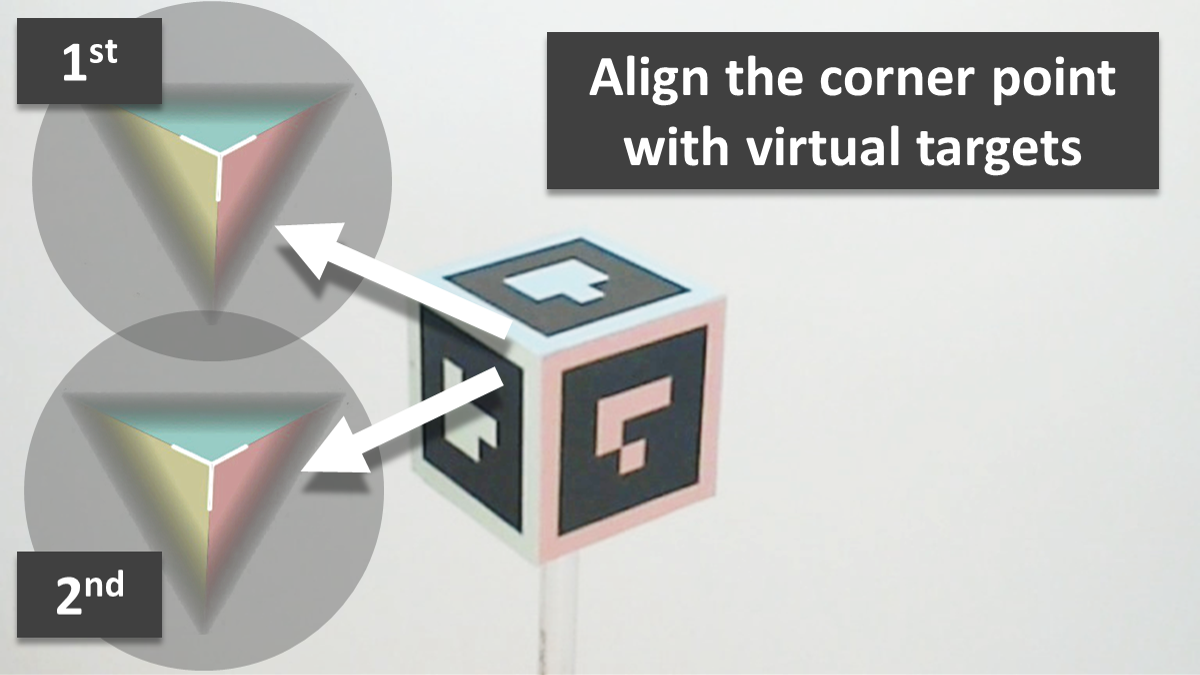}
        \caption{Alignment process with automated voice instructions}
        \label{calib_internal_during}
    \end{subfigure}%
    \hfill
    \begin{subfigure}[t]{0.3\linewidth}
        \centering
        \includegraphics[width=0.95\linewidth]{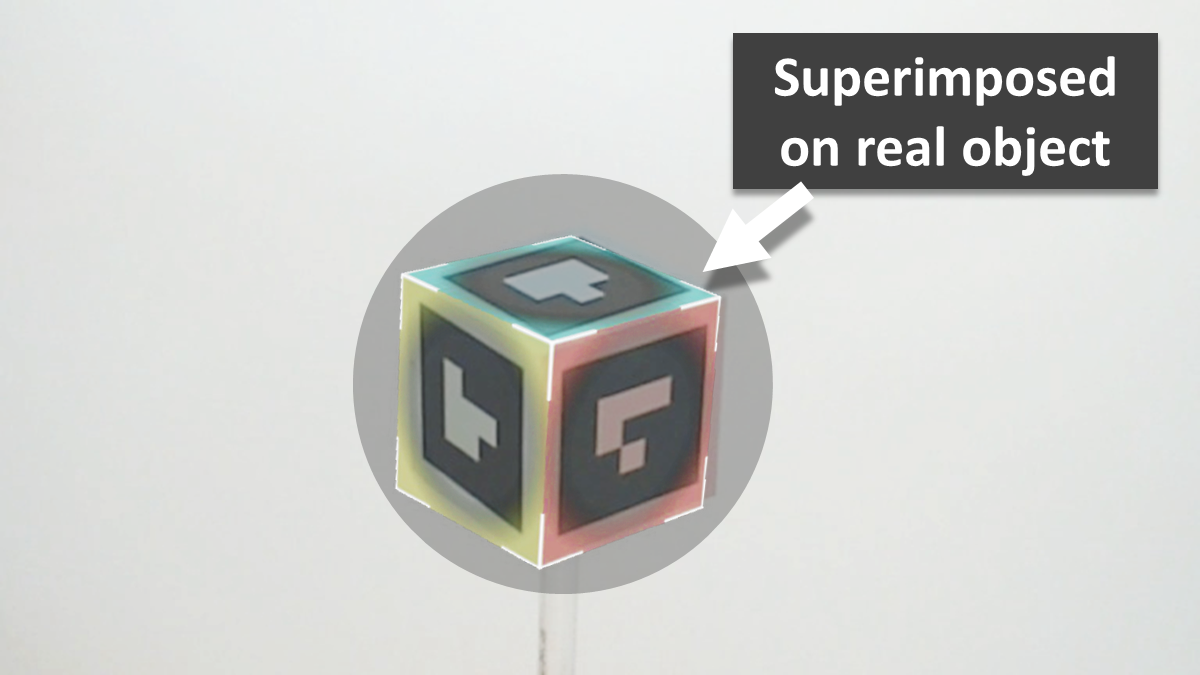}
        \caption{Superimposed cube after calibration}
        \label{calib_internal_after}
    \end{subfigure}
    \caption{Calibration process using head-anchored tracker}
    \label{internal_calib}
\end{figure*}

\begin{figure*}[tbh]
    \begin{subfigure}[t]{0.3\textwidth}
        \centering
        \includegraphics[width=0.98\linewidth]{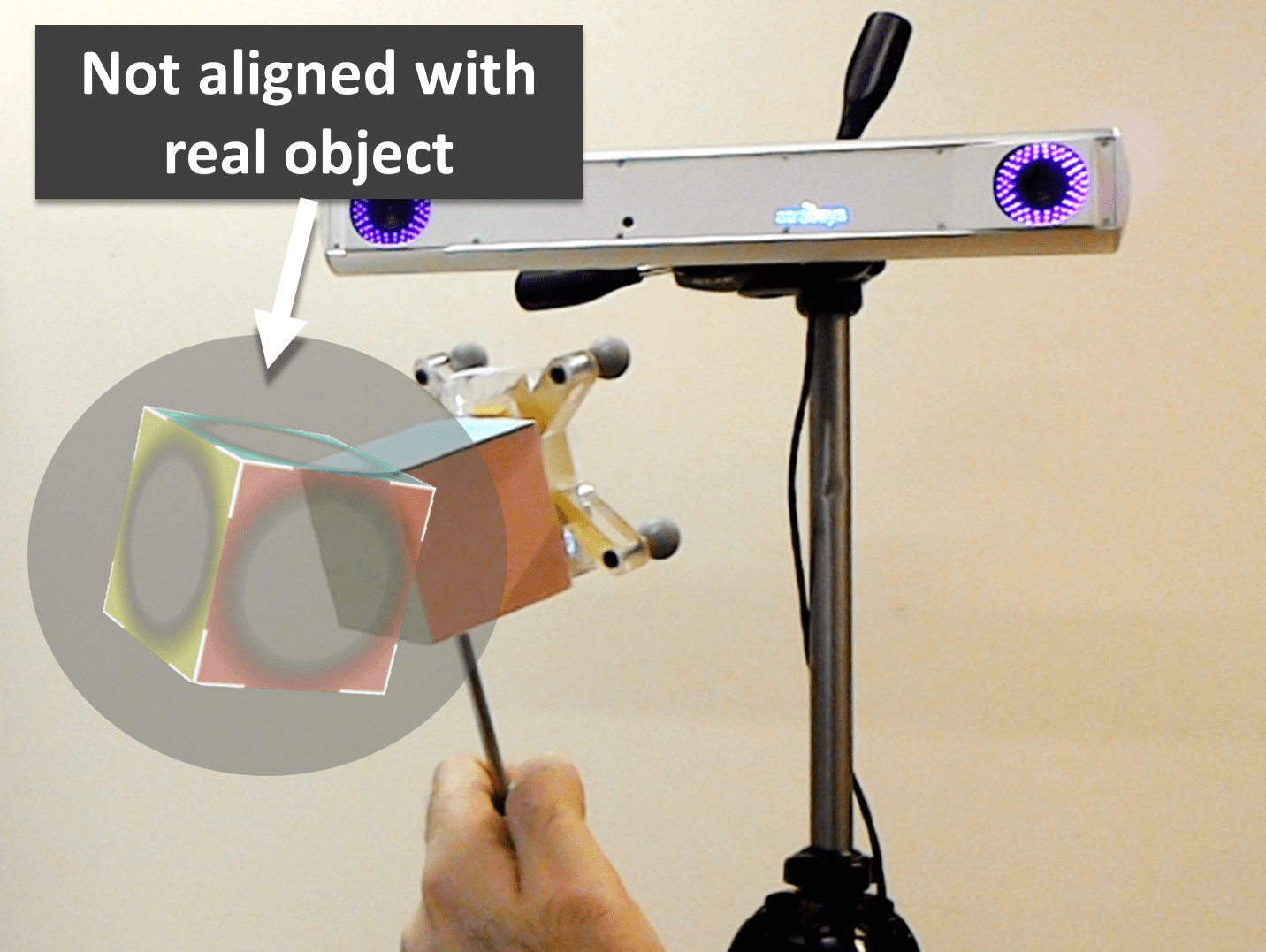}
        \caption{See-through view before calibration}
        \label{calib_external_before}
    \end{subfigure}%
    \hfill     
    \begin{subfigure}[t]{0.3\textwidth}
        \centering
        \includegraphics[width=0.98\linewidth]{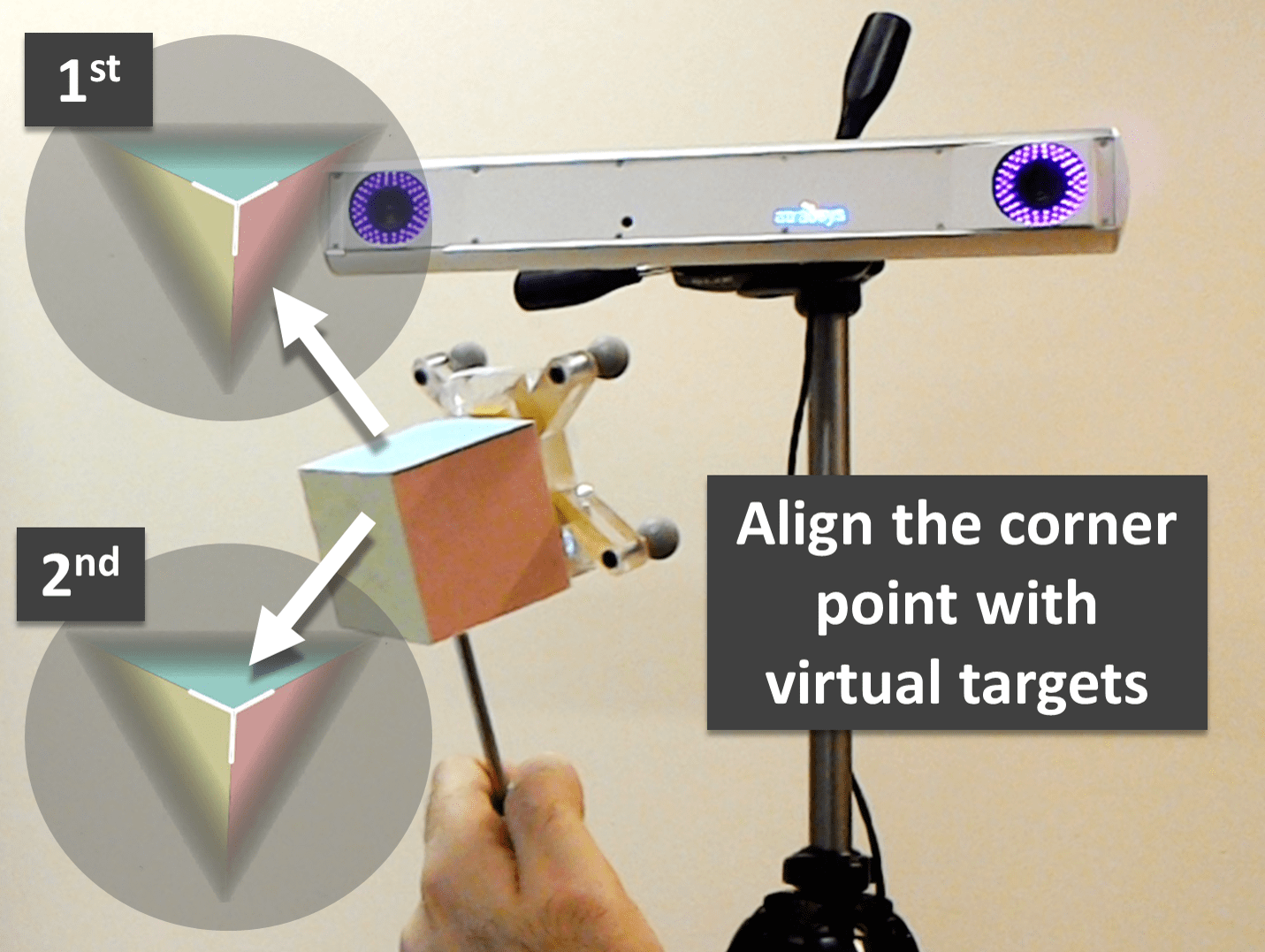}
        \caption{Alignment process with automated voice instructions}
        \label{calib_external_during}
    \end{subfigure}%
    \hfill
    \begin{subfigure}[t]{0.3\textwidth}
        \centering
        \includegraphics[width=0.98\linewidth]{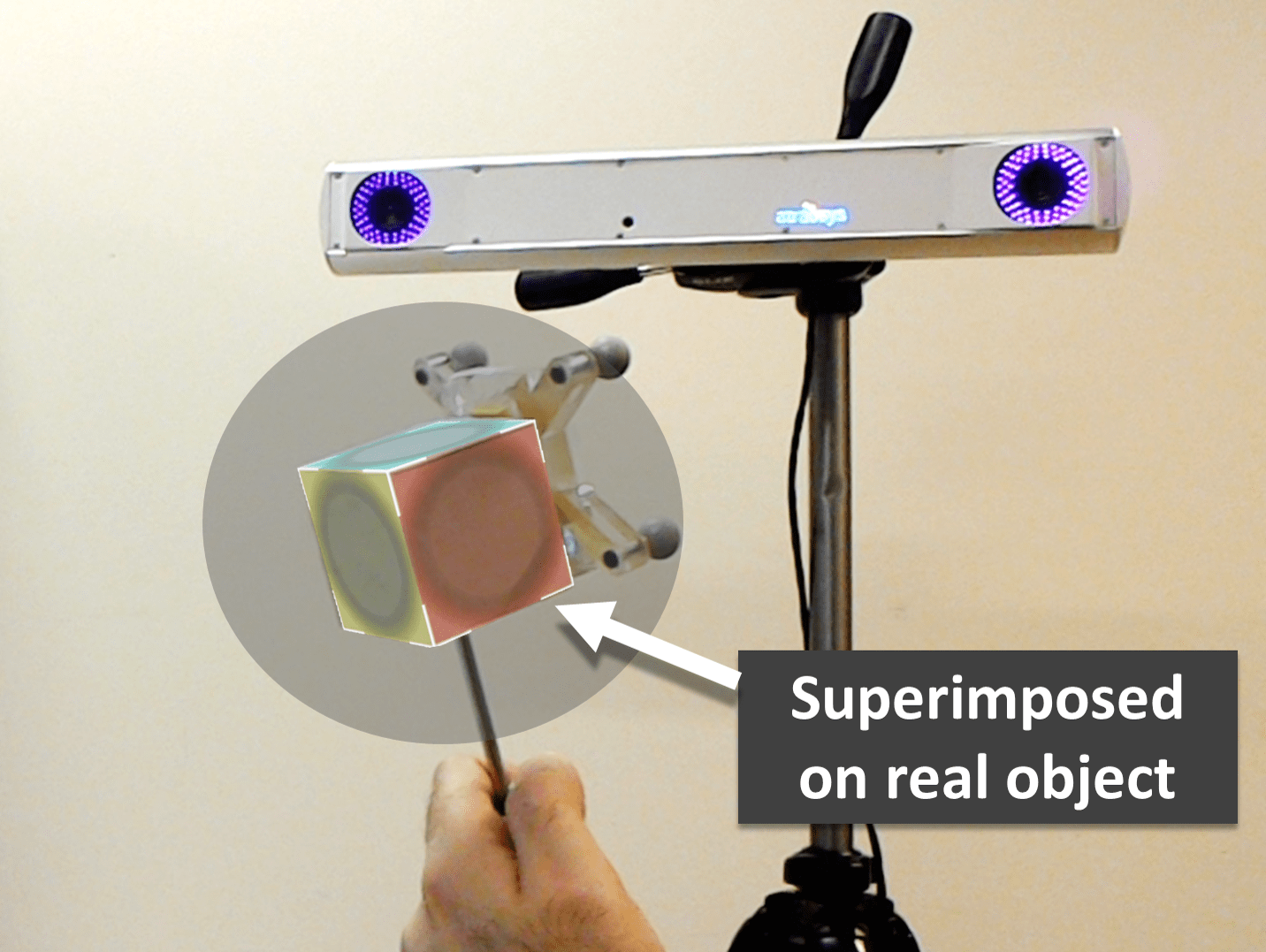}
        \caption{Superimposed cube after calibration}
        \label{calib_external_after}
    \end{subfigure}
    \caption{Calibration process using world-anchored tracker}
    \label{external_calib}
\end{figure*}

\subsection{Standard Method (One Corner Alignment)}
\label{sec:experiment_standard}

The general calibration procedure is the same for each type of tracker.
First, as represented in Figs.~\ref{setup_internal} and~\ref{setup_external}, the user wears the OST-HMD (HoloLens or Moverio BT-300) and is given a real object for alignment. Before calibration, the virtual overlay is not correctly aligned with the tracked cube, as shown in Figs.~\ref{calib_internal_before} and \ref{calib_external_before}. Using automated voice commands, s/he is then instructed to perform the calibration step by step. First, a virtual cube is displayed and the user should try to align only one corner of the cube with its real counterpart in her/his hand (Figs.~\ref{calib_internal_during} and ~\ref{calib_external_during}). Once s/he is satisfied with the alignment, a button is clicked for the confirmation. It should be noted that in this standard method only the corner position is measured for the alignment, however the virtual scene also illustrates a cube with colored faces that matches the real cube. This makes the alignment task more intuitive with the additional depth cue and color similarity. Next, the virtual cube appears in another location in the user's field of view. We try to cover the entire workspace within the reach of the user so that our calibration results are balanced and less biased towards a certain geometrical location. This process continues until 20 points are collected. At this point, the affine, perspective and isometric 3D projection matrices are calculated with their corresponding reprojection errors. The user is able to select from the transformations with different geometrical models, and see the virtual cube in the HMD display environment superimposed on the real cube (Figs.~\ref{calib_internal_after} and ~\ref{calib_external_after}).

\subsection{Multipoint Single 3D Object Method}
This is basically the same as the setup described in the previous section, except the user has to align 5 corners of the cube instead of only one and all of those points are used for the calculations. Again, when the user is satisfied with all the alignments, a button is pressed to confirm and store those points and the virtual cube appears at another position. The user has to repeat this 4 times and the system informs the user of the conclusion of the alignment task.

\subsection{Evaluation}
\label{sec:evaluation}

Evaluating an OST-HMD calibration has always been challenging because only the user wearing it can observe the superimposed objects that result from the calibration. In order to make this process more objective, some studies used a camera instead of the user's eye~\cite{makibuchi2013vision} and performed their measurements on the images captured by that ``eye camera.'' However, using a camera will inevitably affect the depth information.
Measurements of error can be expressed in different ways, such as dimensionless pixels, distance, and visual degrees. Depending on the application, each have their own value.
Here we are using errors in distance since both our world and HMD display coordinate systems use metric Cartesian dimensions. Others also use this metric to report their accuracy~\cite{jun2016calibration}.
In some applications, it is necessary to express accuracy in dimensional units such as millimeters.
For instance, in surgical navigation, two calibrations that both have 5 pixel reprojection error along an axis will have significantly different implications in locating the tumor if that error is equivalent to 5 millimeters or 50 centimeters in the real world. However, in other scenarios where the user's perception is being analyzed, the visual angle or pixel values may play a more important role.  Two evaluation methods are devised to better analyze the accuracy of our proposed calibration.

\noindent\textit{i) Calibrate-and-Test:}
The first evaluation method, which we call calibrate-and-test, is the standard approach where evaluation is performed with additional samples that were not used for the calibration.
Specifically, the user is asked to collect 8 additional samples, and these samples are tested against the calibration calculated with the data sets consisting of the 20 alignments. Reprojection error of the test data is computed based on each of the three transformation matrices (perspective, affine, isometric).
\label{traintest}

\noindent\textit{ii) Double-Cube-Match:}
In this method, a second cube marker is used as an auxiliary reference for objective measurement of the calibration error.
Using the computed transformation from the calibration phase, a virtual cube is displayed in the virtual scene with a predetermined offset of 150 mm with respect to the first marker cube at four different equidistant positions. The user is now asked to align the second real marker cube with the displayed virtual cube as precisely as possible. 

\begin{figure}[!t]
  \centering
    \begin{subfigure}[t]{0.49\linewidth}
        \centering
        \includegraphics[width=\linewidth]{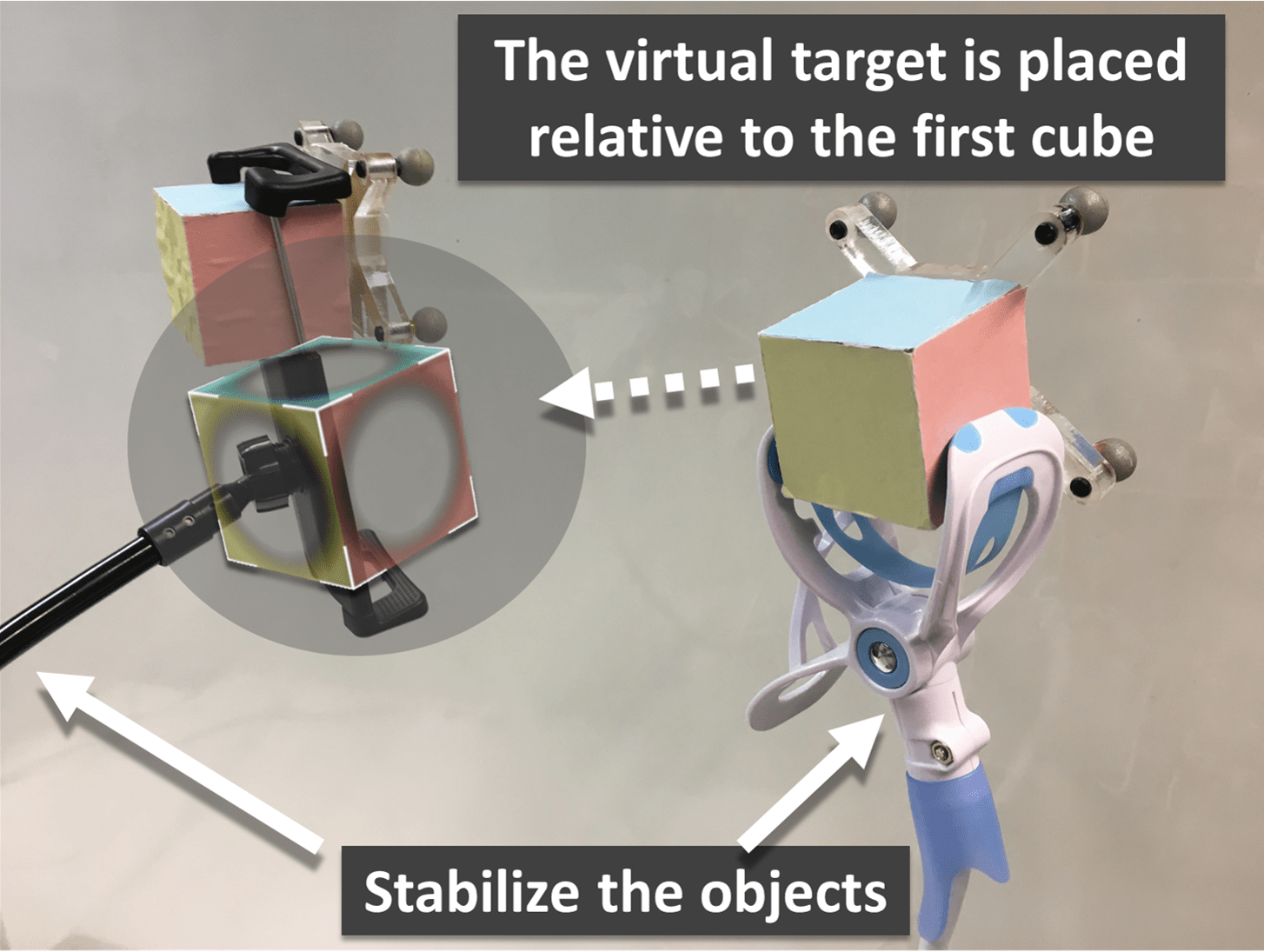}
        \caption{View of the first cube and its virtual counterpart with a 150 mm offset before the alignment}
        \label{eval_external_before}
    \end{subfigure}%
    \hfill
    \begin{subfigure}[t]{0.49\linewidth}
        \centering
        \includegraphics[width=\linewidth]{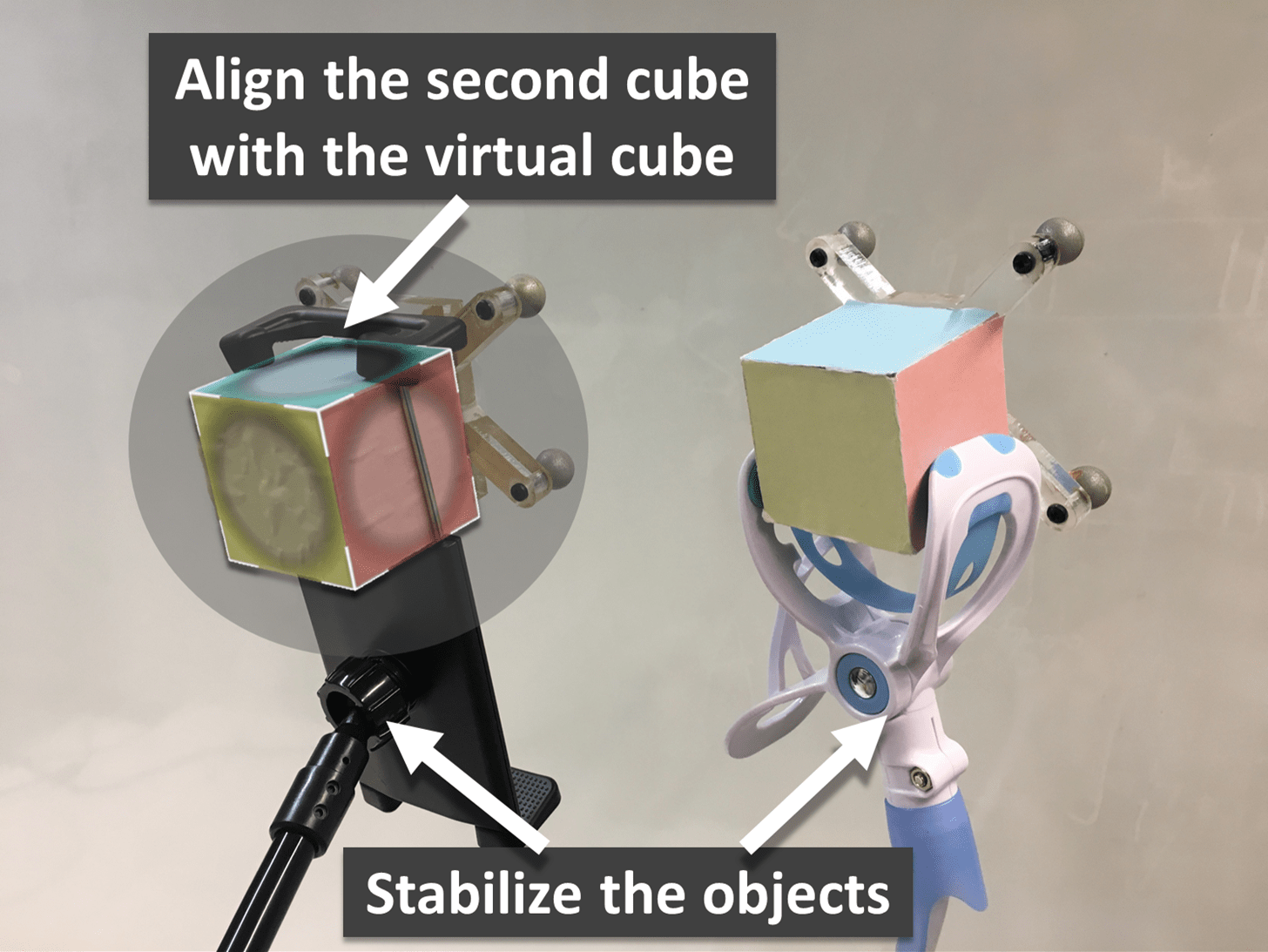}
        \caption{Display after the alignment of the second tracked cube with the first generated virtual cube}
        \label{eval_external_after}
    \end{subfigure}
    \caption{Evaluation process using external measurement system}
    \label{eval_external}
\end{figure}

There are three major sources of error contributing to our observation from this evaluation. In addition to errors generated in the calibration process that we are seeking, there is also error that can be caused by poor alignment. Another error is due to uncertainties in the self localization system of the HMD. To minimize the effect of poor alignment and also eliminate the effects of hand tremor on the alignment in the evaluation process, the tracked cubes are mounted on a flexible stand as shown in Fig.~\ref{eval_external}. With this setup, it is no longer necessary to keep the cube in the user's hand. This allows the user to move around the target and verify the proper alignment from different viewpoints, where distances from the object are no longer limited by the user's arm length. When s/he is satisfied with the alignment, a button is pressed on the keyboard to record the positions of the markers at that instant. The difference between the predetermined offset pose and the observed object pose at the point of the user's confirmation is used as the metric to evaluate the accuracy of the calibration. One might argue that the alignment is still reported by the user, but we assume that as long as the measurements are done by an independent observer (accurate tracking system) to which the user is blind, it serves well for the scope of optical see-through calibration from the perspective of the end user. 

The reported metric in this developed evaluation should be representative of the whole transformation yet intuitive enough that it can be analyzed with similar systems. Therefore, average displacement error $E_{\Delta \X}$ as well as average rotation error using quaternion representation $E_{\Delta\x}$, following the method described in~\cite{markley2007averaging}, are calculated for error analysis.

\section{Results}

The graphical representation of the tracked calibrated cube with its virtual counterpart is illustrated in Figs.~\ref{calib_internal_after} and \ref{calib_external_after}, against the uncalibrated ones in Figs.~\ref{calib_internal_before} and \ref{calib_external_before}. It can be seen that the calibrated one is well aligned, while the uncalibrated one is visibly misaligned. The calibration process was able to superimpose the virtual cube in the correct pose with regard to its real tracked counterpart.

Two users familiar with various HMD systems and calibration techniques performed the calibration with each tracking system in 10 trials followed by the corresponding evaluations. The results and error analysis are presented here.

\begin{figure}[t]
    \begin{subfigure}[t]{0.5\linewidth}
        \centering
        \includegraphics[width=0.96\linewidth]{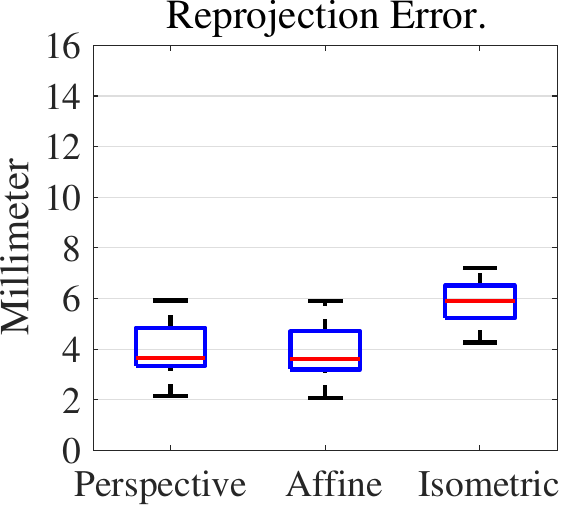}
        \caption{HoloLens}
        \label{calib_internal_measure}
    \end{subfigure}%
    \hfill
    \begin{subfigure}[t]{0.5\linewidth}
        \centering
        \includegraphics[width=0.96\linewidth]{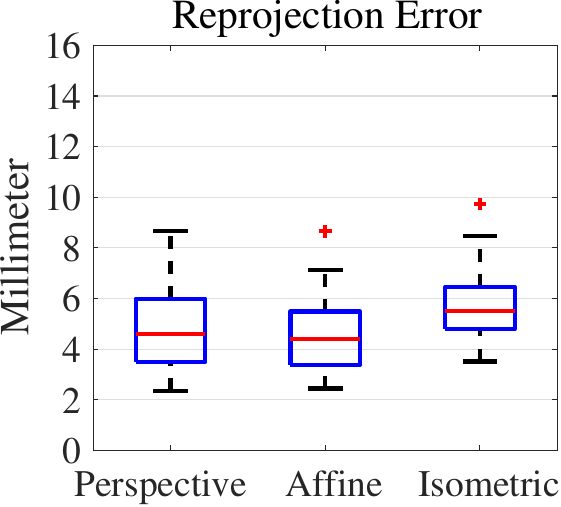}
        \caption{Moverio BT-300}
        \label{calib_internal_measure_Moverio}
    \end{subfigure}
    \caption{Evaluation result of Calibrate-and-Test for the calibration with head-anchored tracker~with two different HMDs, based on different geometrical models.}
    \label{measure_results}
\end{figure}

\subsection{Calibration with Head-Anchored Tracker}

The calibrate-and-test evaluation method is applied to the calibration with the head-anchored tracker, for the three different geometrical models. In each trial (20 trials in total), 20 alignments are used for calibration and 8 alignments are used for testing. To show the flexibility of our proposed method, we performed evaluation on two different OST-HMDs: Microsoft HoloLens and Moverio BT-300.

\subsubsection{Microsoft HoloLens:}

At the calibration stage, the mean and standard deviation of the residue of each geometrical model are: perspective transformation ($3.10 \pm 0.98\,mm$), affine transformation ($3.24 \pm 1.02\,mm$), and isometric transformation ($5.55 \pm 0.67\,mm$). One alignment is identified as an outlier and excluded at this stage. As expected, the calibration residue decreases as the number of parameters in the geometrical model increases.

Fig.~\ref{calib_internal_measure} depicts the reprojection error of the testing data for the HoloLens using perspective, affine and isometric transformation matrices.
The mean and standard deviation of the reprojection error are: perspective ($4.04 \pm 1.04\, mm$), affine ($3.96 \pm 1.06\,mm$), and isometric ($5.86 \pm 0.81\,mm$).
Table~\ref{internal_measure_axis} shows the mean and standard deviation of the reprojection error along different axes for different geometrical models. Here, the $xy$ plane is perpendicular to the user's view and the $z$ axis is parallel to the user's line of sight, indicating the depth of alignment.

\begin{table}[b]
  \caption{Reprojection error along different axes for calibration with head-anchored tracking system for HoloLens}
  \label{internal_measure_axis}
  \begin{center}
    \begin{tabular}{c|cc|cc|cc}
      {} & \multicolumn{2}{c|}{Axis $X$ (mm)} & \multicolumn{2}{c|}{Axis $Y$ (mm)} & \multicolumn{2}{c}{Axis $Z$ (mm)} \\
      Model & mean & std & mean & std & mean & std \\
    \hline
      Perspective & 1.00 & 0.81 & 0.91 & 0.68 & 3.55 & $\,$ 2.62 \Tstrut \\
      Affine &  0.94 & 0.74  & 0.83 & 0.63  & 3.51 & 2.67  \\
      Isometric & 1.82 & 1.08 & 2.05 & 1.36  & 4.58 & 3.31
    \end{tabular}
  \end{center}
\end{table}

\subsubsection{Moverio BT-300:}

At the calibration phase, the mean and standard deviation of the residue of each geometrical model are: perspective transformation ($3.13 \pm 1.10\,mm$), affine transformation ($3.24 \pm 1.14\,mm$), and isometric transformation ($5.07 \pm 1.67\,mm$). One alignment is identified as an outlier and excluded at this stage.

Fig.~\ref{calib_internal_measure_Moverio} depicts the reprojection error of the testing data for the Moverio BT-300 using perspective, affine and isometric transformation matrices. The mean and standard deviation of the reprojection error are: perspective ($4.75 \pm 1.63\, mm$), affine ($4.60 \pm 1.55\,mm$), and isometric ($5.76 \pm 1.57\,mm$).
Table~\ref{internal_measure_axis_Moverio} shows the mean and standard deviation of the reprojection error along different axes for the different geometrical models. As before, the $xy$ plane is perpendicular to the user's view and the $z$ axis is parallel to the user's line of sight.
As described in Section \ref{sec:head-anchored}, the Moverio BT-300 was tested with inaccurate projection matrices and therefore the low reprojection errors achieved with our method serve as yet another indicator of the functionality of the blackbox approach.

\begin{table}[b]
  \caption{Reprojection error along different axes for calibration with head-anchored tracking system for Moverio BT-300}
  \label{internal_measure_axis_Moverio}
  \begin{center}
    \begin{tabular}{c|cc|cc|cc}
      {} & \multicolumn{2}{c|}{Axis $X$ (mm)} & \multicolumn{2}{c|}{Axis $Y$ (mm)} & \multicolumn{2}{c}{Axis $Z$ (mm)} \\
      Model & mean & std & mean & std & mean & std \\
    \hline
      Perspective & 1.11 & 1.07 & 1.18 & 1.13 & 4.07 & $\,$ 3.54 \Tstrut \\
      Affine &  0.96 & 0.90  & 1.07 & 0.97  & 4.02 & 3.49  \\
      Isometric & 2.30 & 1.52 & 1.45 & 0.98  & 4.42 & 3.84
    \end{tabular}
  \end{center}
\end{table}

\subsection{Calibration with World-Anchored Tracker}

Both Calibrate-and-Test and Double-Cube-Match metrics are used to evaluate the calibration of the Microsoft HoloLens with the world-anchored tracking system.

Calibrate-and-Test evaluation of the calibration with the world-anchored tracker is similar to that of the head-anchored tracker. The residue at the calibration stage is as follows: perspective transformation ($3.90 \pm 0.82\,mm$), affine transformation ($4.03 \pm 0.87\,mm$), and isometric transformation ($9.93 \pm 1.47\,mm$). One alignment outlier is identified and excluded. 

\begin{figure}[t]
    \centering
    \begin{subfigure}[t]{0.5\linewidth}      
        \includegraphics[width=0.96\linewidth]{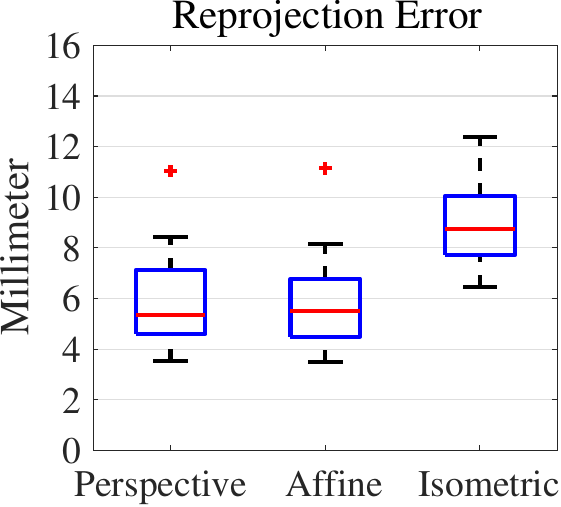} 
    \end{subfigure}
    \caption{Evaluation result of Calibrate-and-Test for the calibration with world-anchored tracker using the HoloLens, based on different geometrical models.}
    \label{calib_external_measure}
\end{figure}

The reprojection error of the calibration results applied on the testing dataset is shown in Fig.~\ref{calib_external_measure}. For perspective transformation, the mean and standard deviation of the reprojection error is $5.88 \pm 1.81\,mm$, while the affine transformation yields an error of $5.83 \pm 1.78\,mm$ and the isometric transformation yields an error of $8.92 \pm 1.60\,mm$. Table~\ref{external_measure_axis} demonstrates the distribution of error along the different axes of the world coordinate system.

\begin{table}[b]
  \caption{Reprojection error along different axes for calibration with world-anchored tracking system for HoloLens}
  \label{external_measure_axis}
  \begin{center}
    \begin{tabular}{c|cc|cc|cc}
      {} & \multicolumn{2}{c|}{Axis $X$ (mm)} & \multicolumn{2}{c|}{Axis $Y$ (mm)} & \multicolumn{2}{c}{Axis $Z$ (mm)} \\
      Model & mean & std & mean & std & mean & std \\
    \hline
      Perspective & 2.47 & 2.04 & 3.01 & 2.49 & 3.20 & $\,$ 3.01 \Tstrut \\
      Affine &  2.44 & 1.98  & 2.98 & 2.52  & 3.21 & 3.01  \\
      Isometric & 3.64 & 2.75 & 6.14 & 3.88  & 3.43 & 2.93
    \end{tabular}
  \end{center}
\end{table}

\begin{figure}[t]
    \begin{subfigure}[t]{0.5\linewidth}
        \centering
        \includegraphics[width=0.96\linewidth]{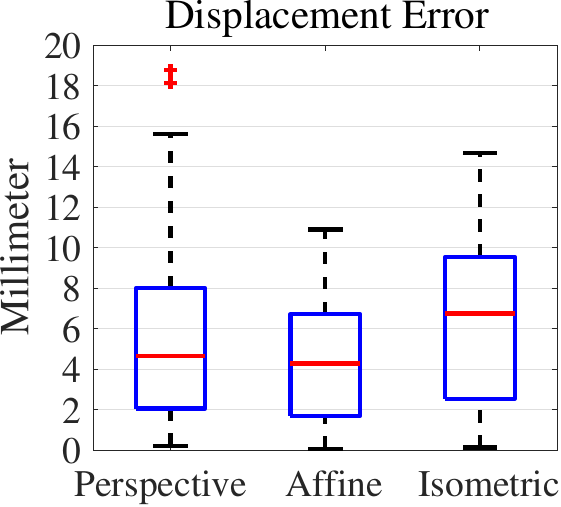}
        \caption{Displacement error $E_{\Delta P}$}
        \label{advanced_position}
    \end{subfigure}%
    \hfill
    \begin{subfigure}[t]{0.5\linewidth}
        \centering
        \includegraphics[width=0.96\linewidth]{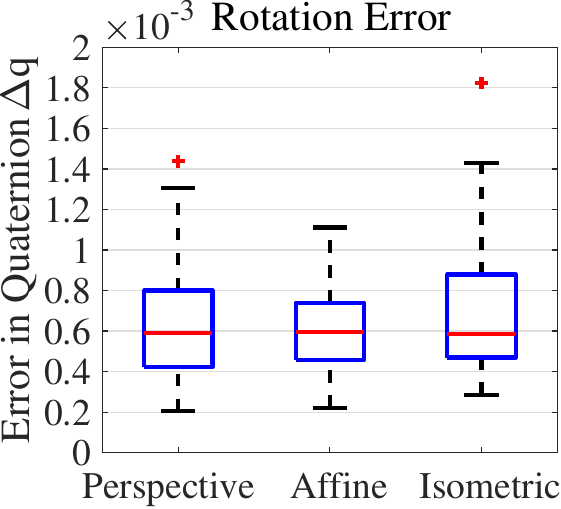}
        \caption{Rotational error $E_{|\Delta\x|}$}
        \label{advanced_rotation}
    \end{subfigure}
    \caption{Evaluation result using Double-Cube-Match method, with world-anchored tracker and HoloLens}
    \label{advanced_results}
\end{figure}

Fig.~\ref{advanced_results} depicts the evaluation results of the Double-Cube-Match metric. Fig.~\ref{advanced_position} shows the distribution of displacement with the three different models.
More specifically, for perspective transformation, the displacement error is $5.47 \pm 4.26\,mm$; for affine transformation, the displacement error is $4.45 \pm 3.00\,mm$, and for isometric transformation, the displacement error is $6.44 \pm 4.15\,mm$. 

Orientation errors in Double-Cube-Match were measured and are shown in Fig.~\ref{advanced_rotation}. Excluding a couple of outliers for the isometric case, the user was able to match the orientation of cubes with a very high accuracy using all calibration matrices. The average error in quaternion for the affine transformation $E_{\Delta\x}$ is given by $(0.999,\,0.005,\,0.002,\,0.007)$ where $\x$ follows $\x = (w,\,x,\,y,\,z)$ representation. Similarly, for the perspective transformation, the quaternion error is $(0.999,\,0.001,\,0.001,\,0.001) $ and $(0.999,\,0.009,\,0.002,\,0.003) $ for the isometric transformation. This excellent accuracy in orientation can be attributed to the fact that the entire virtual cube is visualized in Double-Cube-Match and the user can take multiple viewpoints to adjust the orientation of the real cube using the color similarity and achieve an accurate alignment.

\subsection{Multipoint Alignment with Head-Anchored Tracker}

\begin{figure}[t]
\begin{subfigure}[t]{0.5\linewidth}
    \centering
      \includegraphics[width=0.96\linewidth]{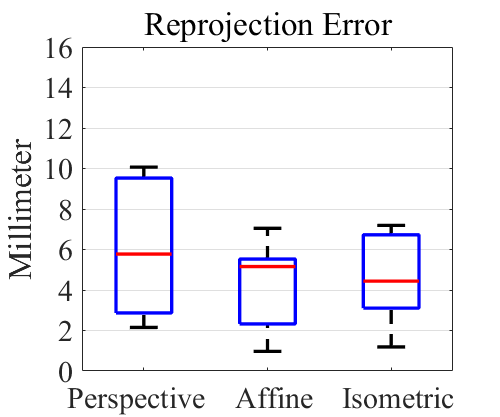}
    \caption{Reprojection error}
    \label{fig:measure_results_multi}
    \end{subfigure}%
    \hfill
    \begin{subfigure}[t]{0.5\linewidth}
        \centering        \includegraphics[width=0.76\linewidth]{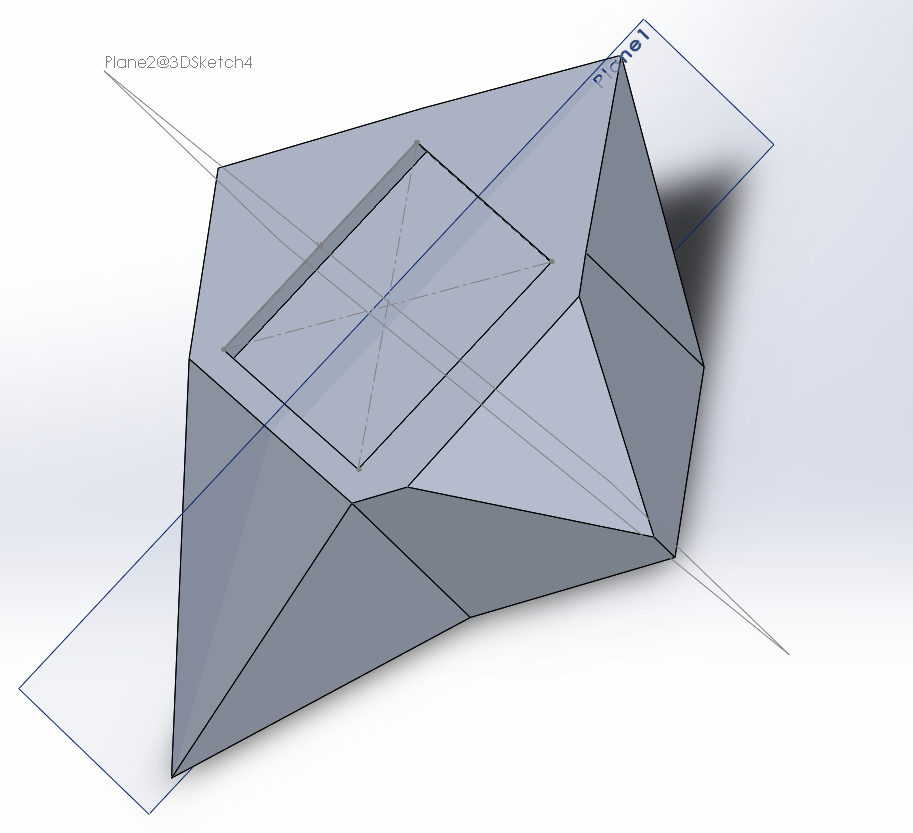}
    \caption{Asymmetrical calibration rig}
    \label{fig:calib_rig}
    \end{subfigure}%
    \caption{Multipoint calibration using single 3D object}
    \label{fig:multipoint}
\end{figure}

The multipoint alignment method was tested with the Microsoft HoloLens.
The reprojection error of the calibration results applied on the testing dataset is shown in Fig.~\ref{fig:measure_results_multi}. For the affine transformation, the mean and standard deviation of the reprojection error is $4.20 \pm 2.20\,mm$, while the perspective transformation yields an error of $6.20 \pm 3.50\,mm$, and the isometric transformation yields an error of $4.60 \pm 2.20\,mm$.

\section{Discussion}

As discussed earlier in Sections~\ref{sec:method_geo} and~\ref{traintest}, different datasets are used for testing our transformation than those used for computing it. This is to ensure that a better result is not simply a better fit but indeed a better representative model.  

The results of the standard (single corner) calibration method, with internal or external tracker, indicate that
at both the calibration and testing stages, the performance of the perspective and affine models are similar, and are both moderately better than the isometric model. In this case, the affine model is the most suitable representation of the transformation between the space of the tracker and the frame of the virtual scene due to the reduced number of unknowns.
With the external tracker, the Double-Cube-Match evaluation method confirms the results of the reprojection error evaluation, showing
that the affine model captures the transformation from the Atracsys tracker to the HoloLens display better than both the perspective model and the isometric model. 
These results are expected since the transformation between two coordinate systems is affine.
Also, the lower accuracy of the perspective model indicates that it is not accurately representing the projection, as it likely overfits the calibration data by adding some distortion that does not exist in the affine model.
The reason that affine performs better than the isometric model can be partly due to scaling that exists between these two coordinate systems.

For the multipoint calibration method, the isometric and affine models provided lower reprojection error for the test data, which suggests that the perspective model overfit the calibration data. It is not surprising that the isometric model performed well in this case, as it is the only transformation that preserves the geometric relationship and dimensions between multiple points on a single object.

Tables~\ref{internal_measure_axis} and~\ref{internal_measure_axis_Moverio}
show that the error of the alignment in depth is still largest among the three directions, for all geometrical models, with the head-anchored (internal) tracking system. Depth alignment is difficult because the visual appearance of varied depth is not as salient as that of the $xy$ plane. Note that a tracking system with a single camera is expected to have lower accuracy in the depth direction.  In addition, considering our tracking system, marker-based \noindent\textit{inside-out} optical tracking studies also show that the tracking result is most vulnerable along the depth direction~\cite{bauer2007tracking,freeman2007method}. Nevertheless, efforts have been made in this work to utilize the depth cue of the 3D immersive display to provide a better depth perception for the users during alignments.
In contrast,
in the case of calibration with the world-anchored tracker, the $x$, $y$, and $z$ axes are parallel to the world coordinate system $\{W\}$ and the user is able to move around and make alignments from different viewing perspectives. Therefore, no axis is associated with the depth direction, as indicated by the consistency of the error values in Table~\ref{external_measure_axis}.

When compared with other recent OST-HMD calibration methods~\cite{jun2016calibration}, our proposed method has a higher accuracy both in the standard and multipoint variation. In addition, their fast method needs at least 10 repetitions to reach $\sim\,10\,mm$ accuracy, while our multipoint single 3D object version needs 4 repetitions to achieve $\sim\,4\,mm$ accuracy. Likewise, our method generates better results when compared with stereo-SPAAM~\cite{jun2016calibration}.

It should also be noted that tracking volumes for these three setups are different. For the head-anchored tracking system with HoloLens, the calibration volume is a frustum (close plane: $110.88\,cm^2$, far plane: $38.88\,cm^2$, depth range: $12\,cm$) due to the limit of field-of-view of the HoloLens~($35^{\circ}$) and the user's arm reach. For the head-anchored tracking system with Moverio BT-300, the calibration volume is a frustum (close plane: $70.58\,cm^2$, far plane: $26.55\,cm^2$, depth range: $12\,cm$) due to the limit of field-of-view of the Moverio BT-300 ($23^{\circ}$) and the user's arm reach. The world-anchored tracking system covers a larger volume ($60\,cm \times 60\,cm \times 60\,cm$). Therefore, the amount of error should be analyzed in its own context; for example in surgical navigation one millimeter may matter, while in a first-person game we may care only about the user's perception and accuracy in depth may play a less important role.

Finally, it is also important to note that despite the considerable efforts and development of other methods for display calibration, to the best of our knowledge, no manufacturer has incorporated a comprehensive calibration process in their system. We believe that this is partly because existing methods are not intuitive enough to be used and often require a tedious task for the users. Presenting this capability, integrated with packages that are widely used by the community including non-experts, allows the adaptation of this indispensable component of user interaction in mixed-reality by a much larger audience. This blackbox approach enables them to use their application on any HMD without worrying about the technical details of each individual system.


\section{Conclusions and Future Work}

In the present study, we proposed a blackbox approach for solving the transformation between the tracking coordinate system and the virtual scene coordinate system. We applied our method for calibration of head-mounted displays using various tracking systems using affine, perspective and isometric transformation models.

We then extended our method and proposed a fast and intuitive multipoint version, in which the user aligns multiple points of the same 3D object in each step at the same time, thereby reducing the number of alignment steps from 20 repetitions to 4. 

Experimental results indicated that the accuracy of our calibration was up to 4 mm, in terms of the average reprojection error, and our more objective evaluation indicated that the average displacement error was almost 4 mm as well.

In addition, we successfully incorporated the self localization and spatial mapping of an OST-HMD in a calibration process, thereby eliminating the need for line of sight from the HMD to the tracker or object.

To address the challenge of subjectivity in evaluation of OST-HMDs, a new evaluation method, namely Double-Cube-Match, was proposed which is less subjective and incorporates an external measurement system as an independent observer.

In the future, we aim to integrate head-anchored and world-anchored tracking systems for HMDs like the HoloLens using sensor fusion, thereby overcoming issues such as occlusion or limited field of view.
In addition, we will adjust our setup to be compatible with different calibration objects. In particular, we note that our use of a cube as the calibration object required us to use different colors on its faces to enable the user to find the correct corner. In an effort to remove this color requirement, we have designed an asymmetrical calibration rig, shown in Fig.~\ref{fig:calib_rig}, that does not have the problem of ambiguity of its corners and can be used in a monochromatic setting.
Finally, we plan to conduct a full user study to investigate issues such as fatigue and user-friendliness of the proposed methods, with comparison of the single point version and the multipoint variation, as well as other existing methods.

\acknowledgments{
We thank Mingkang He, Amir Soltanianzadeh, and Yang Yue for implementing the multipoint calibration component of the application and providing the data and calibration rig shown in Figure~\ref{fig:multipoint}. The authors would also like to thank Alexander Plopski for his helpful comments and feedback.
This work was supported in part by the US Army Medical Research and Materiel Command under Contract W81XWH-15-C-0156, awarded to Juxtopia LLC. The contents should not be construed as an official Department of the Army position, policy or decision unless so designated by other documentation.
}

\bibliographystyle{abbrv-doi}

\bibliography{main}
\end{document}